 final typo etc corrections May 2004

\documentstyle[preprint,aps,epsfig]{revtex}

\begin{document}
\preprint{}
\draft 
\newcommand{\be}{\begin{equation}} 
\newcommand{\ee}{\end{equation}} 
\newcommand{\ba}{\begin{eqnarray}} 
\newcommand{\ea}{\end{eqnarray}} 
\newcommand{\etal}{{\em et al.}}

\title{Extended sudden approximation model for high-energy nucleon removal reactions}

\author{F. Carstoiu$^{1,2}$, E. Sauvan$^{1}$\thanks{Present address: CPPM, 
Marseille, 
France.},  N. A. Orr$^{1}$\thanks{orr@lpccaen.in2p3.fr}, A. Bonaccorso$^{3}$}

\address{$^1$ Laboratoire de Physique Corpusculaire, IN2P3-CNRS, ISMRA et \\ 
Universit\'e de Caen, F-14050 Caen cedex, France \\
$^2$ IFIN-HH, P.O. Box MG-6, 76900 Bucharest-Magurele, Romania \\
$^{3}$ Istituto Nazionale di Fisica Nucleare, Sezione di Pisa, 56127 Pisa, Italy \\}

\maketitle

\begin{abstract}
A model based on the sudden approximation has been developed to describe high
energy 
single nucleon removal reactions. Within this approach, which takes as its
starting point the formalism of Hansen \cite{Anne2}, the nucleon-removal cross 
section and the full 3-dimensional momentum distributions of the core  
fragments 
including absorption, diffraction, Coulomb and nuclear-Coulomb interference 
amplitudes, have been 
calculated. The Coulomb breakup has been treated to all orders for the dipole 
interaction.
The model has been compared to experimental 
data for a range of light, neutron-rich psd-shell nuclei. Good agreement was found 
for both the inclusive cross sections and momentum distributions.
In the case of $^{17}$C, comparison is also made with the results of calculations
using the transfer-to-the-continuum model.  
The calculated 3-dimensional momentum distributions exhibit longitudinal and
transverse momentum components that are strongly coupled by the reaction for $s$-wave states,
whilst no such effect is apparent for $d$-waves.
Incomplete detection of transverse momenta arising from 
limited experimental acceptances thus leads to 
a narrowing of the longitudinal distributions for nuclei with significant $s$-wave 
valence neutron configurations, as confirmed by the data.
Asymmetries in the longitudinal momentum distributions attributed to
diffractive dissociation are also explored.
\end{abstract}

\pacs{PACS number(s): 24.10.-i, 25.60.-t, 25.60.Gc }


\section{Introduction}

The breakup of light exotic nuclei has proven over the last decade to
be a particularly useful tool to study
the structure  of nuclei far from stability (see, for example, \cite{hanse,jim}).  
In the specific case of well
developed single-nucleon halo nuclei, such as $^{11}$Be, much insight has been
gained into the reaction process. 
Early work by Serber
\cite{Serber}, Glauber \cite{Glauber} and Dancoff \cite{Dancoff} (later 
refined by Faldt \cite{fald}), investigating high-energy deuteron 
breakup, demonstrated 
that the relevant
mechanisms governing the reaction are stripping, diffraction and 
Coulomb dissociation. More recently these concepts have been adapted 
by Hansen to describe quantitatively the breakup of the one-neutron halo
nucleus $^{11}$Be \cite{Anne2,Anne1}. 

The basic premise of the model -- dubbed the
``sudden approximation'' -- is that the reaction proceeds by the instantaneous
removal of a nucleon from the projectile without disturbing the
remaining nucleons. This approximation is justified for
reaction times much shorter than the characteristic time for the motion of the
nucleons within the projectile.
A number of 
assumptions make the calculations particularly simple whilst retaining 
the essential physical concepts: (i) the 
incident energy is high
enough so that the intrinsic velocity of the valence nucleon is much smaller
than the projectile velocity; (ii) the projectile and the fragment follow
straight-line 
trajectories; (iii)  
final-state interactions are neglected; (iv) the tail of the valence nucleon 
wave function
is well developed, so reactions involving the valence particle are essentially
surface peaked; (v) the target nucleus can be described by a ``black disk'', so
that the nucleon scattered or absorbed by the target is not observed;
 (vi) there is only one bound state of the system, the
ground state (the completeness of the wave
functions thus allows the transition probabilities to the continuum to be
calculated via 
sum rules);
 and (vii)
only the dominant part (the transverse component) of the momentum transfer
generated by the Coulomb field of the target 
is considered. 

These asumptions are well satisfied in the case of reactions involving 
a well developed
one-neutron halo nucleus such as $^{11}$Be. In this case, where the
$rms$ radius of the halo is some three times larger than that of the
core, only the asymptotic part of the halo wave
function was required (see (iv)) \cite{Anne2,Hansen}.  This leads to analytical
 formulae for the transition 
probabilities in impact parameter space and the longitudinal momentum distributions.
In particular, in evaluating the longitudinal momentum distributions,
Hansen  proposed
that to a good approximation the wave function of the valence nucleon could be
evaluated at the center of the target \cite{Hansen}. This approximation leads to 
the
interesting
result that in the limit of very small binding energies the breakup cross
section factorizes into the free neutron-target cross section and the 
probability
that the neutron is at the center of the target with a longitudinal momentum close
to the incident momentum per nucleon. Thus, the model is intimately related to
the spectator model of Hussein and McVoy \cite{huss}. Whilst this approximation 
is well suited for
the evaluation of longitudinal momentum distributions, it leads to a strong cutoff 
in the
 perpendicular momenta  and is, therefore, less appropriate for the
 evaluation of the momentum distributions in a plane perpendicular to
the beam direction.

 In a recent report, Barranco and
Vigezzi \cite{Barra} have employed a similar approximation  and obtained the 
result that the width of the
perpendicular momentum distribution essentially reflects the target size and 
is
more or less independent of the structure of the projectile ground-state  
wave function. This conclusion does not agree   
 with recent  work we have undertaken whereby the  
 transverse momentum
 distributions were demonstrated to be sensitive to  nuclear structure \cite{nous,sauv01}. 
The sudden approximation model requires a proper normalization 
of the
various transition probabilities, and therefore cannot be extended easily 
to cases where the ground state is dominated by $p$ or $d$ waves. In such cases,
for neutrons 
the asymptotic part of the radial wave function is given by Haenkel
functions\footnote{$h_l^{+}(i\alpha r)$, where $l$ is the angular momentum and $\alpha$ 
is
the decay constant.}, which exhibit a strong singularity at the origin and the
corresponding breakup probabilities cannot be defined properly. Clearly, in 
such a
case, the transition probabilities should be defined in terms of realistic 
wave functions.   

As discussed in our earlier papers \cite{nous,sauv01} and 
elsewhere \cite{hanse,PGHJAT} 
single-nucleon removal reactions appear to be a powerful tool to probe structure
far from stability.
It is therefore highly desirable to test 
and
compare various models of the reaction process in order that spectroscopic information
may be extracted with confidence.
In this spirit we have extended the sudden approximation
model to deal with single-nucleon removal reactions where the ground 
state has an arbitrary structure for which the single-particle degrees of 
freedom
can be disentangled. The  model is
conceptually  simple and almost all observables may be calculated with a 
reasonable numerical
effort even when realistic wave functions are used. 

The paper is organized as follows. Sections II to V are devoted to the basic
formalisms and development of the model. A comparison of the model
calculations and experimental data for single-neutron removal is described in 
Section VI. For comparison the predictions of the transfer-to-the-continuum model \cite{bb,bb2} 
for $^{17}$C are also given.  The effects of finite detection acceptance are 
discussed in Section VII. 
Conclusions are drawn and summarized in Section VIII. Analytical formulae for various
observables are provided in the Appendices.


\section{Basic Formalism}

We assume that the ground state of the projectile $(J^{\pi})$ can 
be  
approximated
by a superposition of configurations of the form  $[I_c^{\pi}\otimes 
nlj]^{J^{\pi}}$, where
$I_c^{\pi}$ denotes the core states and $nlj$ are the quantum numbers
specifying the  single-particle
wave function of the valence nucleon. This is evaluated in a  Woods-Saxon
potential using the 
effective
separation energy, $S^{eff}_n=S_{n}+E_x$ ($E_x$ being the excitation 
energy of 
the core state).
Couplings of core states to the final state and dynamical excitation 
of 
core states in the reaction are neglected. In this approximation the reaction 
can populate a given core
state only to the extent that there is a nonzero spectroscopic factor 
$C^2S(I^\pi_c,nlj)$ in the
projectile ground state. When more than one configuration contributes to 
a given core state,
the total cross section for single-nucleon removal is written, 
following refs. \cite{Navin,Tost}, as an 
incoherent superposition of single particle cross sections,
\be
\sigma_{-1n}(I^\pi_c)=\sum_{nlj} C^2S(I^\pi_c,nlj) \; 
\sigma_{sp}(nlj,S^{eff}_n)
\label{eq1}
\ee
The total inclusive single-nucleon removal cross section 
($\sigma_{-1n}^{sudd}$) is then the sum over the cross sections to all 
core states. A similar relation holds for the momentum distributions. 
 The term $\sigma_{sp}$ is the cross section
for the removal of a nucleon from  a single-particle state with total angular
momentum $j$, 
\be
\psi_{jm}=R_{jl}(r)\sum_{m_l,m_s}C^{l~~ {{1}\over{2}}~~ j}_{m_l m_s m}Y_{lm_l}
(\hat r)\chi_{{{1}\over{2}}m_s}(\sigma).
\label{eq2}
\ee
evaluated with the  effective nucleon separation
energy $S_{eff}$ defined above.
We consider only spin independent transition operators and therefore all
formulae are much simpler with the wave function,
\be
\psi_0(\vec r)=R_l(r)Y_{lm}(\hat r),
\label{eq3}
\ee
and the normalization,
\be
\int d\vec r\vert\psi_0(\vec r)\vert^2=1.
\label{eq4}
\ee

Certain cross sections may be expressed in a simpler form in a coordinate system 
traveling with
the beam. The impact parameter with
components in the direction connecting the centre of the core with that of the target 
is denoted by $\vec b$. 
Recoil effects are of the order $O(\frac{1}{A_p})$, where $A_p$ is the 
projectile mass number and are neglected here. The
situation is illustrated in Fig.~\ref{fig:figsudd}. After the interaction 
with the target a part of
the wave function is removed. 
 At this stage the specific form of the wound ($(w)$) is not important. The 
removed part of the wave function is then,

\begin{displaymath}
\delta\psi(\vec r)=\left \{\begin{array}{ll}\psi_0(x,y,z)&  \textrm{if} ~
(x,y,z)\in $(w)$\\
                   0& \textrm{otherwise.}\\
\end{array}\right.
\label{eq5}
\end{displaymath}

The complement ($\bar\psi$) is defined by the following orthogonal decomposition 
of
the wave function,
\ba
\psi_0=\bar\psi+\delta\psi\\
\int d\vec r\bar\psi^{*}\delta\psi=\int d\vec r\bar\psi\delta\psi^{*}=0
\label{eq6}
\ea
If the nucleon is absorbed, the wave function of the system is localized
to $\delta\psi$ at the instant of collision and this is then the state of the
remaining core fragment. Consequently, the momentum distribution is given by
the square of the Fourier transform of $\delta\psi$. The stripping (absorption)
probability is given by the volume integral of the wound,
\be
P_a(b)=\int d\vec r\vert\delta\psi\vert^{2}
\label{eq7}
\ee
As mentioned in ref. \cite{Anne2}, ``absorption'' is defined here in the context 
of the black disk model.  The wound complement is normalized as,
\be
\int d\vec r\vert\bar\psi\vert^{2}=1-P_a
\label{eq8}
\ee
The wave function after collision with the target is approximated by,
\be
\psi (\vec r)=e^{i\vec q\vec r}(\psi_0-\delta\psi)=e^{i\vec q\vec r}\bar\psi
\label{eq9}
\ee
where $\vec q$ is the momentum transfer to the valence nucleon defined by 
Eq.~(A.4) in  Appendix A. The sudden transfer of a momentum $\vec q$ attaches
a phase $e^{i\vec q\vec r}$ to the wave function. The physical meaning of the
phase is clarified in Section V. This wave function has the 
normalization
,
\be
\int d\vec r \psi^{*}(\vec r)\psi(\vec r)=\int d\vec r \bar \psi^{*}(\vec r)
\bar\psi(\vec r)=1-P_a
\label{eq10}
\ee
Note that $\psi$ contains elastic as well as inelastic (breakup) states. The 
elastic content is given by the overlap with the ground state,
\be
\gamma_{el}=\int d\vec r \psi\psi_0^{*}(\vec r)
\label{eq11}
\ee
The wave function orthogonal to the ground state is,
\be
\psi_1(\vec r)=\psi(\vec r)-\gamma_{el}\psi_0(\vec r)
\label{eq12}
\ee
Now we are ready to construct the wave function for the decaying state 
($\psi_{d}(\vec r)$),
\be
\psi_{d}(\vec r)=\psi_0(\vec r)-\delta\psi(\vec r)-\gamma_{el}e^{-i\vec q
\vec r}\psi_0(\vec r)\equiv \bar\psi-\gamma_{el}e^{-i\vec q\vec r}\psi_0\equiv
 e^{-i\vec q\vec r}\psi_1(\vec r)
\label{eq13}
\ee

As the state which is decaying contains only square integrable functions it is 
legitimate to speak about the norm of this wave function,
\be
\int d\vec r \vert \psi_d(\vec r)\vert^2=\int d\vec r \vert \psi_1(\vec r)
\vert^2=1-P_a-\vert\gamma_{el}\vert^2
\label{eq14}
\ee
Clearly, all information concerning the absorption of the valence nucleon is 
contained in the wave function of the wound ($\delta\psi$), whilst $\psi_d$ will furnish 
information on the elastic breakup. Let us calculate the Fourier transform of 
the decaying state,
\be
\widetilde{\psi_d}(\vec k)=\frac{1}{(2\pi)^{3/2}}\int d\vec r e^{-i\vec k\vec r}
\psi_d(\vec r)
\label{eq15}
\ee
The differential cross section in momentum space is given by,
\be
\frac{d\sigma}{d \vec k}=\int 2\pi b db\vert\widetilde{\psi_d}(\vec k)\vert^2
\label{eq16}
\ee
and the total cross section by,
\be
\sigma=\int 2\pi b db\int d\vec k\vert\widetilde{\psi_d}(\vec k)\vert ^2=\int 
2\pi b db(1-P_a(b)-\vert\gamma_{el}(b)\vert^2)
\label{eq17}
\ee
Now it is clear that the quantity, 
\be
P_{el}(b)=1-P_a(b)-\vert\gamma_{el}(b)\vert^2
\label{eq18}
\ee
should be interpreted as the probability in the impact parameter representation
 of the elastic breakup process.  Relation (\ref{eq18}) shows that the sudden 
approximation model accounts for stripping ($P_a$), elastic breakup or 
dissociation ($P_{el}$) and nuclear and Coulomb elastic scattering 
($\gamma_{el}$).
Other inelastic processes such as core absorption, simultaneous 
absorption of the valence neutron and the core \cite{Hencken} or ``damped
breakup'' \cite{Nego} are not included 
in this model. We also note that the closure relation (\ref{eq18}) illustrates the need 
to use 
realistic wave functions in order to obtain properly normalized breakup 
probabilities.


\section{Breakup probabilities}

In this section we discuss in detail the calculation of the breakup 
probabilities using two specific forms for the wound $(w)$. Using the same
notations as in ref. \cite{Anne2} the planar cutoff is defined as,
\begin{displaymath}
(w):(x,y,z)\in\mathbf{R}^{3},~x\geq b_1=b-R_t
\label{eq19}
\end{displaymath}
The second, the so called cylindrical wound is defined  for arbitrary $z$ as,
\begin{displaymath}
(w):(x,y,z)\in\mathbf{R}^{3},~(x-b)^2+y^2\leq R_t
\label{eq20}
\end{displaymath}
where $R_t$ denotes the target radius.
For  clarity it is useful to define the (normalized) valence density by 
averaging over projections of the angular momentum $m$,
\be
\rho_{val}(r)=\frac{1}{\hat l^2}\sum_m\vert\psi_0\vert^2=\frac{1}{4\pi}R_l^2(r)
\label{eq21}
\ee
 Writing $\vec r\equiv (x,y,z)\equiv(\vec s,z)$, we define 
also one and two-dimensional projections of this 
density as,
\be
\tilde\rho(x)=\int dy dz\rho_{val}=2\pi\int_x^{\infty}r dr \rho_{val}(r)
\label{eq22}
\ee
\be
\hat\rho(s)=\int dz\rho_{val}(r)=2\int_s^{\infty}\frac{r dr}{\sqrt{r^2-s^2}}
\rho_{val}(r)
\label{eq23}
\ee
The singularity in (\ref{eq23}) is weak and can be integrated by parts. 
Without loss of generality one can assume reflection symmetry,
 $\rho_{val}(-r)=\rho_{val}(r)$, then the mapping $x\rightarrow\tilde\rho(x)$ 
is a reflection symmetric homomorphism,~$\tilde\rho(-x)=\tilde\rho(x)$
 and $2\int_0^{\infty}dx\tilde\rho(x)=1$. Since the valence density
$\rho_{val}$ is normalized to one, the density $\hat\rho$ also satisfies
the closure relation $\int d\vec s \hat\rho(s)=1$.

\subsection{Elastic probability}
According to Eq.~(11) the elastic probability  defined by,
\be
\gamma_{el}=\int d\vec r\psi_0^*(\vec r)e^{i\vec q\vec r}(\psi_0-\delta\psi)
\equiv\gamma_C-\gamma_{C+N}
\label{eq24}
\ee
decomposes into a Coulomb and a nuclear+Coulomb amplitude. It is easy to show 
that,
\ba
\gamma_C=2\int_0^{\infty}dx \cos qx\tilde\rho(x)\\
\gamma_{C+N}=2\pi\int_{b_1}^{\infty}dx e^{iqx}\tilde\rho(x)
\label{eq25}
\ea
For a cylindrical wound the Coulomb and nuclear amplitude is somewhat more 
complicated
\be
\gamma_{C+N}=\int_0^{R_t}s ds\int_0^{2\pi}d\phi e^{iq(b+s\cos\phi)}\hat\rho
(\sqrt{b^2+s^2+2bs\cos\phi})
\label{eq26}
\ee
Note that $\gamma_{C+N}$ is a complex function and depends on the specific 
form of the 
wound, whilst the Coulomb amplitude is real and wound independent. Moreover 
$\lim_{q\rightarrow 0}\gamma_{C}=1$ which shows that in the absence of a Coulomb 
field or at very large impact parameters the Coulomb amplitude is 
unity. However, the long range Coulomb interaction make this convergence very
slow.
In the case of a  well developed neutron halo, the wave function varies 
little over the wound region and one can replace in (\ref{eq26}) $\hat\rho$ by 
some average value $\hat\rho_0$ which leads to,
\be
\vert\gamma_{C+N}\vert^2=\hat\rho_0^2\frac{4\pi^2R_t^2}{q^2}J_1^2(qR_t)
\label{eq27}
\ee
in analogy with the Fraunhofer diffraction pattern produced by an
absorbing disk. This approximation illustrates the diffraction 
content 
of $\gamma$ \cite{Barra}. In practice 
this amplitude is evaluated using the exact equations (\ref{eq25}) and 
(\ref{eq26}).


\subsection{Absorption/stripping probability}

The stripping probability is evaluated using Eq.~(7),
\be
P_a(b)=\int_{(w)}dx dy dz \rho_{val}(r)
\label{eq28}
\ee
which leads after some simple manipulations to,
\be
P_a(b)=\int_{b_1}^{\infty}dx \tilde\rho(x)
\label{eq29}
\ee
in the case of a planar cutoff and
,
\be
P_a(b)=\int_{0}^{R_t}sds\int_{0}^{2\pi}d\phi\hat\rho(\sqrt{b^2+s^2+2bs\cos\phi})
\label{eq30}
\ee
for a cylindrical wound. Independent of the form of the wound we have $P_a(0)=1/2$
and in the limit $q\rightarrow 0$,~
$\gamma_{C+N}(b)=P_a(b)$ and $P_{el}(b)=P_a(b)-P_a^2(b)$. The total cross 
sections
 for stripping and diffraction are
obtained by integration of the
 above probabilities over the impact parameter with the volume element 
$2\pi bdb$. The above relations imply that for a light target, where the Coulomb
component is negligibly small,  
$\sigma_{str}\approx\sigma_{diff}$.  In general, however 
$\sigma_{str}>\sigma_{diff}$
 in agreement with the original formulation of Glauber \cite{Glauber}.


\section{Momentum distributions}

As final state interactions are neglected, the momentum distributions in the 
coordinate system traveling with the beam are given by the square of the 
Fourier transform (Eq.~(15)) of the wave function Eq.~(13). The three components 
in Eq.~(13)
lead to an amplitude of the form, $A_{lm}(\vec
k)=A_{0}+A_{\delta}+A_{C+N}$, where $A_{0}$ is the unperturbed (intrinsic)
amplitude. Using the same notations as 
in ref. \cite{Anne2} one has for angular momentum $(lm)$,
\be
A_{0,lm}(\vec k)=\frac{4\pi}{(2\pi)^{3/2}}i^lY_{lm}(\hat k)\int_0^{\infty}r^2 
dr j_l(kr)R_l(r)
\label{eq31}
\ee
and 
\be
A_{C+N,lm}(\vec k)=-\gamma_{el}(b)\frac{4\pi}{(2\pi)^{3/2}}i^lY_{lm}(\hat k_q)
\int_0^{\infty}r^2 dr j_l(k_qr)R_l(r)
\label{eq32}
\ee 
where $\vec k_q=(k_x+q,k_y,k_z)$.
The calculation of the amplitude $A_\delta$ is more involved. The difficulty 
arises from the angular part of the 
wave function. The simplest way to proceed is to express the spherical harmonics in 
Cartesian coordinates 
(see for example ref. \cite{Edmonds}). For the planar cutoff case we need to 
evaluate integrals of the form,

\be
I_{pst}(\vec k)=\int_{b_1}^{\infty}dx x^p e^{-ik_{x}x}\int_{-\infty}^{\infty}dy
y^s e^{-ik_{y}y}\int_{-\infty}^{\infty}dzz^t e^{-ik_{z}z}R_l(r)/r^l
\label{eq33}
\ee
where $p,s,t$ are positive integers. In the simplest case, $p=s=t=0$, 
we have,

\be
I_{000}(\vec k)=2\pi\int_{b_1}^{\infty}dxe^{-ik_{x}x}\int_0^{\infty}udu
J_0(k_{\perp}u)R_l(\sqrt{x^2+u^2}/r^l 
\label{eq34}
\ee
where $J_0$ is the cylindrical Bessel function  and $k_{\perp}=
\sqrt{k_y^2+k_z^2}$. For any other 
combination of $p,s,t$, the corresponding integral is 
obtained by the appropriate 
parametric differentiation with respect 
to $k_x,k_y,k_z$. For an $l=0$ wave function the result is,
\be
\Re A_{\delta}(\vec k)=-\frac{1}{\sqrt{2}(2\pi)^2}\int_{b_1}^{\infty}dx \cos 
k_x x\int_0^{\infty}u du J_0(k_{\perp}u)R_0(r)
\label{eq35}
\ee
\be
\Im A_{\delta}(\vec k)= \frac{1}{\sqrt{2}(2\pi)^2}\int_{b_1}^{\infty}dx \sin
 k_x x\int_0^{\infty}u du J_0(k_{\perp}u)R_0(r)
\label{eq36}
\ee
with $r=\sqrt{u^2+x^2}$.
The corresponding expressions for $p$ ($l=1$) and $d$ ($l=2$) states are more 
involved and are 
detailed in Appendix B. 

Once the amplitudes have been computed, the differential cross section 
is obtained by averaging over the
magnetic projections and the impact parameter,
\be
\frac{d\sigma}{d\vec k}=\frac{1}{\hat l^2}\sum_m\int d\vec b\vert A_{lm}(k_x,k_
y,k_z)\vert^2
\label{eq46}
\ee
It can be seen that the reaction mechanism modifies substantially the 
momentum content selected by the reaction. For example, for $l=0$ the 
unperturbed amplitude Eq.~(\ref{eq31}) is spherical and real and the amplitude 
selected by stripping is asymmetric and complex. These effects will be 
discussed in detail in the next section.
It should be noted that all amplitudes tend to zero for $b\rightarrow\infty$ 
as is evident from Eqs. (\ref{eq35})
and (\ref{eq36}).  Furthermore the amplitudes (\ref{eq31}) and (\ref{eq32}) become 
identical for large $b$ as in the limit
$b\rightarrow\infty$, $\gamma_{el}=1$, $\vec k_q=\vec k$ and the two amplitudes
essentially cancel each other.
Asymptotically, the amplitude for diffraction is, therefore, essentially the Fourier
transform of the wound.
 Both stripping and diffraction
amplitudes behave for large $k_x$ as $1/k_x$. The reason for this is of course 
that the leading term of the Fourier
transform of the step function (characteristic of the black disk approximation 
for the S-matrix ) behaves like
$1/k_x$. This property is independent of the asymptotic behavior of the 
wave function and merely characterizes the
sharp edge of the target. 

In more realistic reaction 
models, such as  the Glauber model 
with the S-matrix generated by an optical potential \cite{Nego,nous,sauv01}, 
the 
absorption
 evolves smoothly from 0 to 1 and the
amplitudes fall off faster. One may also note that the differential cross 
section 
 (\ref{eq46}) is calculated in the
intrinsic reference system  for a particular position of the target.
 In this system the diffraction
amplitude is not symmetric with respect to $k_x$ since the Coulomb field pushes
 the core in one direction.
The amplitudes, however, are symmetric to rotations in the $(x,y)$ plane. This is 
obviously not an observable symmetry. In order
to obtain observable $(p_{\parallel},p_{\perp})$ distributions one should 
average over all directions in the $(x,y)$
plane as illustrated in Fig.~\ref{fig:figsudd2}. 
In practical terms, the physical momentum distributions are obtained from  
the following unitary transformation,
\be
\frac{d\sigma}{d \vec p}=\frac{1}{2\pi}\int d\beta\frac{d\sigma}{d\vec
k}(p_x\cos\beta-p_y\sin\beta,p_x\sin\beta+p_y\cos\beta,p_z),
\label{eq47}
\ee
 where the  angle $\beta$ is defined in Fig.~\ref{fig:figsudd2}. 
 Equation (\ref{eq47}) gives the  3-dimensional 
momentum distribution in the laboratory system. It contains all the physics of 
the process and together with (\ref{eq46})
 provide a means to study the effects of finite detection acceptances and 
other experimental effects on the momentum
distributions. Unfortunately the corresponding calculations are very time 
consuming. In practice we have
combined 
competitive Gauss-Legendre
numerical integration (up to  order 98 in the approximating polynomial) with
 Monte Carlo simulations in the last step
(Eq.~(\ref{eq47})) which took into account all experimental broadening effects:  
angular dispersion in the secondary beam, angular and energy straggling
 in 
the target and detector resolution \cite{sauv01}. The Lorentz broadening is  
also included, 
\ba
E_B=\gamma(E_A-\vec \beta \cdot \vec p_A),\\
\vec p_B=\vec p_A+\gamma \vec \beta (\frac{\gamma \vec \beta \cdot \vec
p_A}{\gamma+1}-E_A),
\label{eq48}
\ea
where the reference system A is traveling with the beam velocity $\vec \beta$
and B is the laboratory system.

If the acceptance in the plane perpendicular to the beam direction
is infinite, then it is possible to obtain much simpler formulae for the
longitudinal momentum distributions. These are detailed in Appendix 
\ref{ppar_formulae}.


\section{Coulomb dissociation in the sudden approximation}

 As already remarked in ref. \cite{Anne2}, the Coulomb 
excitation
calculation based on the (nonperturbative) sudden approximation is consistent 
with  perturbation
theory \cite{Bertu} in the sense that it contains its leading term as a limit
valid for low $Z_cZ_t$ (core and target charge) and high velocities. This 
approximation is no longer 
valid for large $Z_cZ_t$  and is thus limited to low Z targets. Moreover, 
in this model the dependence on the projectile energy is  given 
by  the Coulomb contribution alone via the momentum transfer $\vec q$. In more 
sophisticated models of the Glauber type (see, for example, \cite{sauv01} and references therein), 
there is an additional 
energy dependence through the S-matrix elements of the nucleon-target and 
core-target interaction.  This dependence, however, 
is expected to be weak in a high energy regime.

In this section we wish to clarify the Coulomb dissociation effects present 
in the model. In this respect we make contact with the more elaborate calculations of 
ref. \cite{jerome}. This model solves 
the
time dependent three-body problem assuming that the core moves along a classical
(straight line) trajectory $\vec R(t)=\vec b+vt\hat z$, whilst the valence nucleon
 is subject to the interaction,
\be
V_2(\vec r,t)=V_{nt}(\vec r+\vec R(t))+V_{dip}(\vec r,t),
\label{eqff1}
\ee
where the first term is the neutron-target interaction and the second is a
dipole approximation for the ``shake-off interaction''. If only the Coulomb
shake-off is taken into account then for a valence neutron,
\be
V_{dip}(\vec r,t)=\frac{Z_c Z_t e^2}{A_p}~~\frac{\vec r.\vec R(t)}{R^3(t)}
\label{eqff2}
\ee
In the general case one uses the effective dipole charge defined in Appendix A. 
Note that Eq.~(\ref{eqff2}) embodies a pure recoil effect ($\sim 1/A_p$) which retains both
the transverse and longitudinal momentum transfers to the neutron arising from the
projectile-target Coulomb interaction.

The transfer-to-the-continuum model breakup amplitude takes the form,
\be
g_{lm}(\vec k,\vec b)=\frac{1}{i\hbar}\int_{-\infty}^{\infty} dt
<\phi^f\vert V_2(\vec r,t)\vert \phi_{lm}^i>
\label{eqff3}
\ee 
where $\phi^{i(f)}$ are initial (final) state wave functions and $lm$ are, as
before, single-particle quantum numbers carried by the ground state. In the 
eikonal approximation, we have
\be
g_{lm}(\vec k,\vec b)=\frac{1}{i\hbar}\int d \vec r \int dt e^{-i\vec k\vec
r+i\omega t}e^{\frac{1}{i\hbar}\int_t^{\infty}dt'V_2(\vec r,t')}V_2(\vec
r,t)\phi_{lm}(\vec r)\equiv <\vec k\vert I(\omega)\vert lm>
\label{eqff4}
\ee
where $I(\omega)$ is the time integral and $\hbar\omega=\epsilon_k-\epsilon_0$
is the excitation energy. Pure Coulomb effects in the dipole approximation are
obtained by swiching off the neutron-target nuclear interaction $V_{nt}$,
\be
I_{C}(\omega)=\frac{1}{i\hbar}\int_{-\infty}^{\infty} dt e^{i\omega
t}e^{\frac{1}{i\hbar}\int_t^{\infty}dt'V_{dip}(t')}V_{dip}(t).
\label{eqff5}
\ee
In the sudden approximation (sa)  the
excitation energy is negligibly small ($\omega\approx 0$).
 The time integration is thus trivial and we find,
\be
I_{C}^{sa}=e^{-i\chi_C}-1
\label{eqff6}
\ee
with, 
\be
\chi_C=\frac{1}{\hbar}\int_{-\infty}^{\infty}dtV_{dip}(t)=\frac{1}{\hbar}\int_{-
\infty}^{\infty}dt\frac{Z_cZ_te^2}{A_p}~\frac{\vec
s\vec b+zvt}{(b^2+v^2t^2)^{3/2}}=\frac{2Z_cZ_te^2}{A_p}~\frac{\vec s\vec
b}{\hbar v b^2}
\label{eqff7}
\ee
where the neutron position vector relative to the core has been decomposed into
transverse and longitudinal components ($\vec s,z$). Comparison with Eq.~(A4) in
Appendix A shows that,
\be
e^{i\chi_c}\equiv e^{i\vec q\vec r}
\label{eqff8}
\ee
and the complete equivalence of our amplitude $A_{C+N}$ (Eq.~31) with the
corresponding sudden approximation amplitude in the eikonal TC model. Therefore,
the amplitude $A_{C+N}$ contains breakup to all orders in the dipole shake-off
Coulomb interaction at zero excitation energy. It is appropriate for evaluation
of Coulomb effects for light targets. For heavy targets, the radial integral 
in Eq.~(\ref{eq32}) is difficult to evaluate numerically.   
We further stress
that the third term in the wave function (Eq.~(13)) $\gamma_{el}e^{i\vec q\vec 
r}\psi_0$
is nothing other than the high-energy eikonal approximation to the scattering
wave function where in this particular case the eikonal phase due to the dipole
(shake-off) Coulomb interaction is calculated along the unperturbed (straight
line) classical trajectory.


\section{Discussion}

As detailed in refs. \cite{nous,sauv01} we have previously undertaken an experimental 
study of 
high-energy one-neutron removal reactions on a range of $psd$-shell nuclei.
In the present section the inclusive cross sections, longitudinal 
and transverse
momentum distributions  obtained  for reactions on a carbon
target are compared to the results of calculations using the model
developed in the preceeding sections.

As in our earlier work \cite{nous,sauv01} the 
spectroscopic amplitudes entering in Eq.~(\ref{eq1}) have been calculated 
with the aid of the
shell-model code OXBASH\cite{oxbash}.  Where known, the experimentally 
established spin-parity
($J^{\pi}$) assignments and core excitation energies have been used. In all other
cases shell model predictions were employed. The carbon target radius was
fixed to $R_t=1.15(12)^{1/3}$. The core radii have been evaluated with a liquid
drop formula \cite{devisme},
$$
R_c=x_1 A_c^{1/3}(1+x_2 A_c^{-2/3}+x_3 A_c^{-4/3})
$$
with $x_1=1.17$, $x_2=1.225$ and $x_3=-0.115$. The single-particle wave
functions were obtained by solving the Schr\"{o}dinger equation for a
Woods-Saxon (WS) potential including central and spin-orbit terms. The depth of 
the
central potential was adjusted to reproduce the known effective neutron binding 
energy ($S_n^{eff}=S_n+E_x$). The
potential radius was taken to be equal to the core radius and the diffusivity 
was
fixed to $a_{WS}=0.6$ fm\footnote{A fine tuning in the range $0.5-0.65$ fm
 could, in principle, be made for each nucleus to improve the cross section with
 respect to the measured value.}.
  The minimum impact
parameter was defined by $b_{min}=R_c+R_t$ and the maximum impact parameter was
fixed to be $b_{max}=50$ fm which ensured that there were no spurious effects from
the cancellation of the amplitudes of Eqs. (\ref{eq31}-\ref{eq32}). Using the 
upper adiabatic limit
suggested by Hansen \cite{Anne2} produced only minor changes in the cross
sections. We further assumed that the core ground and excited states have the same
density distributions and, as such, the same Woods-Saxon geometry was employed
for all core states of the projectile.

The cross sections calculated within the planar cutoff approximation are 
displayed
in Fig.~\ref{fig:sefsud}. The stripping and diffraction
(including Coulomb breakup) components are also presented. A good
overall agreement with the experimental cross sections is observed. 
Keeping all the parameters
fixed as defined above, the results obtained with a cylindrical wound are shown in
Fig.~\ref{fig:sefsud2}. This calculation systematically underestimates the
experimental value by a factor of two.  This is a 
pure geometrical
effect resulting essentially from the fact that the cylindrical wound
underestimates the real extension of the interaction region.  As such a diffuse edge 
cylindrical wound would probably be
more appropriate.

The particular form of the wound, however, has essentially no influence on the 
shape of
momentum distributions. Calculations in the planar cutoff approximation are
displayed in Fig.~\ref{fig:syssudd}. Monte Carlo filtering of the calculated
distributions including
Lorentz broadening, energy straggling in the target, the emittance of the
beam and detector resolutions have been performed and the resulting
distributions, normalized to the  peaks of distributions are displayed by the solid 
lines in
Fig.~\ref{fig:syssudd}. In addition to the excellent overall agreement, we note
that the distributions are somewhat better reproduced 
by the sudden 
approximation
model as compared to the Glauber calculations employed in our earlier work
\cite{nous,sauv01}, as illustrated in
Figs \ref{fig:exsud} and \ref{fig:px_sudd_glau}.
 This is primarly attributable to the
strong damping of the high momentum components in the wave function by the sharp
cutoff approximation. 

As mentioned earlier, the momentum
distributions are first calculated in the projectile rest frame
which is determined by a particular relative core-target configuration.
The observable momentum distributions are then obtained by the unitary 
transformation of
Eq.~(\ref{eq47}). The effect of this transformation is displayed in Fig.
\ref{fig:s3d_kxpx} for the the $s$-wave valence neutron in the ground state of 
$^{15}$C (S$_n$=1.2 MeV). In the rest frame, the $k_x$ and $k_y$ distributions have
markedly different shapes as the Coulomb shake-off imparts momentum in only 
one
direction ($x$). After averaging over all directions (Eq.~\ref{eq47}) the $p_x$ 
and $p_y$
distributions become identical in agreement with what is observed 
experimentally.
The  asymmetry induced by the Coulomb interaction in the rest frame,
translates into a small broadening effect in the laboratory frame.

Further insight into the role played by the reaction mechanism on the momentum
distribution is explored in Fig.~\ref{fig:s3d_int}. Here momentum
distributions in the transparent limit of the Serber model\cite{Serber} are 
compared with the planar
cutoff calculations in the rest frame for $s$ and $d$-wave valence neutrons. 
As already noted
by Hansen \cite{Hansen}, the longitudinal components ($k_z$) become narrower,
irrespective of the angular momentum carried by the wave function. For the
transverse 
component ($k_x$), the effect is more complex. A broadening 
effect is observed for the
$s$-state, whilst the shape of the distribution is completely changed for a 
$d$-state.
 The overall effect after averaging  (Eq.~\ref{eq47}) is a broadening of
the transverse distributions and a narrowing of the longitudinal one.
These conclusions are consistent with the measurements of ref. \cite{sauv01}
whereby the transverse distributions were found to be sytematically broader than
the longitudinal ones. We note that this is in contradiction to the analysis of
Sagawa and Takigawa, whereby (for $s$-wave states) the transverse distribution
is narrowed by the absorptive cutoff induced by the reaction process
\cite{sag90}. 
The calculated transverse momentum distributions, after inclusion of the
experimental effects, are compared to the data for selected nuclei
in Fig.~\ref{fig:px_sudd}\footnote{Owing to the very time consuming nature of
these calculations, only $p_x$ distributions for four nuclei with ground state
structures representative of those measured in ref. \cite{sauv01} have been
computed.}. The finite angular acceptances of the spectrometer
and detector resolution introduce a smooth cutoff in the high momentum 
tails of the  
distributions.  As in the case of the longitudinal momentum
distributions, the shape and width of transverse distributions show a direct
dependence on the projectile 
structure, and are not simply a reflection of the target size as suggested by
Barranco and Vigezzi\cite{Barra}.

As a further example we display in Fig.~\ref{fig:spectro_px} 
the results of calculations
for $^{17}$C for the three possible ground-state
spin-parity assignments.  The results are very similar 
to those obtained in our earlier work using an extended 
Glauber-type model (Fig.~19 of ref. \cite{sauv01}),
which suggested a spin-parity assignment of 3/2$^+$ arising
from a dominantly $^{16}$C($2^+_1$ $\otimes$ $\nu d_{5/2}$)  
configuration.  As noted in refs \cite{nous,sauv01}, this is in 
line with the direct observations of coincident 1.76~MeV 
$\gamma$-rays by Maddalena {\em et al.} \cite{mada} and
more recently by Datta Pramanik {\em et al.} \cite{datta}. 

 For completness, the perpendicular momentum
distributions ($p_{\perp}=\sqrt{p_x^2+p_y^2}$) have also been reconstructed
\cite{these} from the data of refs. \cite{nous,sauv01}. The calculations 
for selected nuclei are compared, after Monte Carlo filtering of the experimental
effects to the data in Fig.~\ref{fig:pper_sudd}
where very good agreement is again found.

As intimated earlier, one of the goals of the present work was to explore
a reaction model other than the Glauber-type approach which has been
the principle means to utilising high-energy nucleon removal as a spectroscopic
tool.  In this spirit we have also performed calculations for the reaction 
of $^{17}$C on a carbon target using the Transfer-to-the-Continuum Model 
(TCM) developed by Bonaccorso and Brink \cite{bb,bb2,bb4}, which has been 
employed to describe with some success single-neutron removal from beams
of $^{34,35}$Si and $^{37}$S \cite{Enders}.  
As described in ref. \cite{bb4}, the TCM formalism includes a more complete dynamical
treatment of the motion of the removed nucleon than Glauber/eikonal models.  
In particular, spin-coupling between the
initial and final states may be included and 
can result in asymmetric momentum distributions.

The TCM calculations presented here (Fig.~\ref{fig:TCM}) were performed as outlined in 
ref. \cite{bb4} (Eqs. (2.1-2.4)).
 The neutron-target optical
potential was taken from ref. \cite{dg} and the strong absorption radius, used in the
parametrization of the core survival probability, was fixed to be 6.8~fm.  
The $^{16}$C 
core longitudinal momentum distribution was obtained from that of the neutron 
by momentum conservation.  As may be seen in Fig.~\ref{fig:TCM}, the results
assuming the $^{17}$C ground-state structure outlined above 
(J$^{\pi}_{gs}$=3/2$^+$)
are in excellent agreement
with both the data and the calculations using the sudden approximation.


\section{Detection acceptance effects}

Following the first measurements of core-fragment longitudinal-momentum
distributions \cite{nao92}, the influence on the observed lineshapes of limited
angular or transverse momentum acceptances has been discussed
\cite{rii93,kel95,kel97,nao95,bazin}. These discussions have, however, been
based on the assumption first introduced by Riisager of a
3-dimensional Lorenzian momentum distribution \cite{rii93}. More recently
it has been conjectured that the reaction 
mechanism itself causes the core-fragment momentum components
to decouple \cite{Hansen}. 
As such, incomplete detection in the plane ($x,y$)
 perpendicular to the beam direction 
 would have no  
influence on the measured longitudinal momentum distribution.  

The data set presented in refs. \cite{nous,sauv01} provides a good opportunity 
to 
investigate such an effect at a quantitative level, as the full 3-dimensional 
momentum distribution of the core-fragment have been acquired. In the off-line 
analysis, 
the angular acceptance of the 
spectrometer was reduced from the full acceptance of $\pm$2$^\circ$ 
to 
$\pm$1.5, 1, 0.5, 0.25$^\circ$. These limits correspond to transverse momenta
$p_x$ and $p_y$ 
of around  $\pm$200, 150, 100, 50 and 25~MeV/$c$. Of the 23 nuclei studied in refs. 
\cite{nous,sauv01} the restricted acceptances lead to a narrowing of the 
longitudinal momentum 
distributions in 5 cases --- $^{14}$B, $^{15,16}$C and $^{24,25}$F. In 
the case of $^{16}$C the narrowing reaches some 25 \% for the most limited
acceptance (Fig.~\ref{fig:cut_angle_width} and \ref{fig:cut_angle16C}).
Interestingly these nuclei have in common a large $s$-wave component in the ground-state 
wave function. For other nuclei, with ground
states dominated by  
 $d$-wave valence neutron configurations, no significant reduction in the widths of 
 the longitudinal momentum distribution was observed.

In the sudden approximation model the full 
3-dimensional core-fragment momentum distribution can be evaluated and
projected onto the desired direction after integrating over the corresponding
acceptances.
For the four nuclei selected earlier as examples, the acceptances applied to the
data have also been applied to the calculated distributions.  
The narrowing with the reduced transverse acceptances of the longitudinal momentum
distributions is well reproduced for  $^{14}$B 
and $^{15,16}$C (Fig.~\ref{fig:cut_angle_width}).  In the case of $^{19}$N, 
experimentally the width diminishes by only some 5\%
as the acceptances are reduced. This trend is well reproduced by the
calculated distributions\footnote{Note that the axis displaying the FWHM in 
Fig.~\ref{fig:cut_angle_width} has been expanded and the widths are in fact
reproduced to within 5\% or better}.  We note that, as intimated above, 
the $^{19}$N ground state is dominated by a $d$-wave valence neutron
configuration.

These calculations indicate that the momentum components are, contrary to the
suggestion of Hansen \cite{Hansen} strongly correlated and 
reduced acceptances in, for example, the transverse direction will  
affect the longitudinal momentum distribution. 
This effect is most evident for $s$-wave 
valence neutron configurations (e.g. $^{14}$B and $^{15,16}$C). The longitudinal
momentum distributions for each acceptance are compared in 
Fig.~\ref{fig:cut_angle16C} for the case of $^{16}$C 
with the calculated distributions and very good
agreement is found.  

For the distributions derived using the full acceptances of the
spectrometer, the experimental distributions are somewhat asymmetric and exhibit low
momentum tails.  This effect is not reproduced by the present model and, as noted 
in our earlier
paper \cite{sauv01}, almost certainly arises in the case of weakly bound
systems as a result of a strong coupling to continuum in diffractive
dissociation.  Such a process has been succesfully modelled by Tostevin \etal
\cite{cdcc} within the coupled discretized continuum chanels (CCDC) formalism, 
where
 the redistribution of relative energy to the internal excitation of
the projectile is treated exactly. As is evident from 
Fig.~\ref{fig:cut_angle16C}, as the angular acceptances are progressively reduced
the asymmetry becomes less pronounced as the contribution from diffraction
decreases, an effect also observed by Tostevin \etal \cite{cdcc}.

Finally, the case of $^{19}$N is displayed in
Fig.\ref{fig:cut_angle19N}. As noted above, within the statistical precision of
the present measurements the width of the longitudinal momentum distribution
remains unchanged with decreasing acceptances.  This supports the model prediction
that for $d$-states the different momentum components are effectively decoupled.
Interestingly the asymmetry in the momentum distribution appears to persists even for
very limited transverse acceptances a feature which cannot be easily explained 
by diffractive processes.  


\section{Conclusions}

The sudden approximation approach for the description of high-energy, 
single-neutron breakup of
halo nuclei, originally introduced by Hansen \cite{Anne2}, 
 has been extended to include realistic  
wave functions and to incorporate shell model spectroscopic
amplitudes. The theory is based
on the strong absorption  description of the core-target and neutron-target
interactions. Applied to single-neutron 
removal reactions, the model allows for the calculation of the full 
3-dimensional 
momentum 
distribution of the core-fragment including stripping, diffraction and
 Coulomb dissociation mechanisms. The Coulomb/nuclear interference is 
also taken into account. As in the case of our Glauber-type calculations 
\cite{nous,sauv01} the
observation that the transverse momentum distributions are systematically
broader than the longitudinal distributions is reproduced by the present
calculations.
 
Calculations were performed for comparison with measurements of
inclusive cross sections and longitudinal and transverse
momentum distributions for a series of some 23 neutron-rich $p-sd$ shell nuclei
 \cite{nous,sauv01}. Suprisingly, for such a relatively simple model, very good 
agreement with the measured cross sections
and 
momentum distributions was found. Indeed, the momentum distributions were
somewhat better reproduced by the present model than the more sophisticated
Glauber-type calculations \cite{nous,sauv01}.

Importantly, the effect of limited detection angular acceptances on 
the longitudinal 
momentum distributions could be investigated. A significant reduction of 
the widths was observed experimentally for nuclei with ground states
dominated by $s$-wave valence neutron configurations. Little or no reduction
was observed, however, 
 for nuclei with dominant
$d$-wave valence neutron components. This effect was well reproduced by  
the present model calculations and is believed to arise 
as a consequence of the correlations between the momentum 
components in the 3-dimensional momentum distribution of the 
core-fragment following the reaction.


\acknowledgements

We are grateful to Gregers Hansen for stimulating discussions. 
One of  the authors (F.C.) acknowledges the  financial support from IN2P3
including that received within  
the  framework of the IFIN-HH$/$IN2P3  convention. F.C. also acknowledges
the kind hospitality shown to him by the staff of LPC-Caen 
during the preparation of this work. Finally we wish to thank our colleagues who
were involved in acquiring data described in refs.\cite{nous,sauv01} for
allowing us to use it here.


\appendix
\section{}

In this appendix we derive a convenient expression for the classical momentum
transfer using straight line trajectories and retaining only the dipole 
component of  the Coulomb field. The projectile consists of core ($m_c,Z_c$)
 and
a cluster ($m_x,Z_x$) moving in the $z$-direction with velocity $v$. The impact 
parameter $b$ with respect to the target is measured in the $x$-direction.
The dipole effective charge  is defined by \cite{Typel} as,

\be
Z_{eff}^{(1)}={{Z_c m_x-Z_x m_c}\over{m_c+m_x}}
\ee

The time dependent electric field due to the target charge $Z_t$ is given by,
\begin{displaymath}
\vec E(t)=\frac{\gamma Z_t e}{(b^{2}+\gamma^{2} v^{2} t^{2})^{3/2}}\left
(\begin{array}{l}b\\0\\vt\end{array}\right)
\end{displaymath}
here $\gamma$ is the Lorentz contraction factor. The classical momentum
transfer is then,
\be
\Delta\vec p=\int_{-\infty}^{\infty}dt e Z_{eff}^{(1)}\vec E(t)
\ee
One sees immediately that $E_y=0\Rightarrow \Delta p_y=0$ and $\Delta p_z=0$
from parity consideration. It follows that the momentum transfer has only one
component in the x-direction given by,
\be
\Delta p_{x}=\gamma Z_t
Z^{(1)}_{eff} e^{2} b
\int_{-\infty}^{\infty}
\frac{dt}
{(b^{2}+\gamma^{2}v^{2}t^{2})^{3/2}}=\frac{2 Z_t Z_{eff}^{(1)} e^{2}}{bv}
\ee                                      
The momentum transfer imparted by the cluster is
,
\be
\vec q=\frac{2 Z_t Z_{eff}^{(1)} e^{2}}{\hbar c \beta}\frac{\vec b}{b^2}
\ee
Momentum conservation gives then the core momentum.
When the cluster is a neutron Eq.~(A4) is identical to Eq.~(10) of ref. 
\cite{Anne2}.

\section{}

Explicit expressions for $A_\delta$ amplitudes (Section IV) are detailed in 
this appendix, for $l=$0, 1 and 2 wave functions. In all expressions the 
notation  
 $r=\sqrt{u^2+x^2}$, is used.

For $l=0$  one has,
\be
\Re A_{\delta}(\vec k)=-\frac{1}{\sqrt{2}(2\pi)^2}\int_{b_1}^{\infty}dx \cos 
k_x x\int_0^{\infty}u du J_0(k_{\perp}u)R_0(r)
\label{eq35b}
\ee
\be
\Im A_{\delta}(\vec k)= \frac{1}{\sqrt{2}(2\pi)^2}\int_{b_1}^{\infty}dx \sin
 k_x x\int_0^{\infty}u du J_0(k_{\perp}u)R_0(r)
\label{eq36b}
\ee

For $l=1$ and $m=\pm 1$,
\be
A_{\delta 1m}=C_{1m}(I_1+I_{2m})
\label{eq37}
\ee
with $C_{1m}=\frac{m\sqrt{3}}{2(2\pi)^2}$ and,
\ba
\Re I_1=2\pi\int_{b_1}^{\infty}x dx \cos k_x x\int_0^{\infty}uduJ_0(k_{\perp}u)
R_1(r)/r\\
\Im I_1=-2\pi\int_{b_1}^{\infty}x dx \sin k_x x\int_0^{\infty}uduJ_0(k_{\perp}u
)R_1(r)/r
\label{eq38}
\ea
and
,
\ba
\Re I_{2m}=2\pi m\frac{k_y}{k_{\perp}}\int_{b_1}^{\infty} dx \cos k_x x\int_0^{
\infty}u^2duJ_1(k_{\perp}u)R_1(r)/r\\
\Im I_{2m}=-2\pi m\frac{k_y}{k_{\perp}}\int_{b_1}^{\infty} dx \sin k_x x\int_0^
{\infty}u^2duJ_1(k_{\perp}u)R_1(r)/r
\label{eq39}
\ea
For $l=1~m=0$,
\be
A_{\delta 10}=\frac{1}{(2\pi)^{3/2}}\frac{1}{2}\sqrt{\frac{3}{\pi}}I_0
\label{eq40}
\ee
with
,
\be
I_0=-2\pi\frac{k_z}{k_{\perp}}\int_{b_1}^{\infty}dx e^{-ik_x x}\int_0^{\infty}u
^2duJ_1(k_{\perp}u)R_1(r)/r
\label{eq41}
\ee
For $l=2~m=\pm 2$,
\be
A_{\delta 2m}=\frac{\sqrt{15}}{8\pi}(I_1+I_2+I_{3m})
\label{eq42}
\ee
\ba
\Re I_1=-\int_{b_1}^{\infty}x^2 dx \cos k_x x\int_0^{\infty}uduJ_0(k_{\perp}u)R
_2(r)/r^2\\
\Im I_1=-\int_{b_1}^{\infty}x^2 dx \sin k_x x\int_0^{\infty}uduJ_0(k_{\perp}u)R
_2(r)/r^2\\
\Re I_2=\frac{k_z^2}{k_{\perp}^3}\int_{b_1}^{\infty} dx \cos k_x
x\int_0^{\infty
}u^2duJ_1(k_{\perp}u)R_2(r)/r^2\nonumber\\
+\frac{1}{2}\frac{k_y^2}{k_{\perp}^2}\int_{b_1}^{\infty} dx \cos k_x x\int_0^{\
infty}u^3du(J_0(k_{\perp}u)-J_2(k_{\perp}u))R_2(r)/r^2\\
\Im I_2=-\frac{k_z^2}{k_{\perp}^3}\int_{b_1}^{\infty} dx \sin k_x x\int_0^{\inf
ty}u^2duJ_1(k_{\perp}u)R_2(r)/r^2\nonumber\\
-\frac{1}{2}\frac{k_y^2}{k_{\perp}^2}\int_{b_1}^{\infty} dx \sin k_x x\int_0^{\
infty}u^3du(J_0(k_{\perp}u)-J_2(k_{\perp}u))R_2(r)/r^2\\
\Re I_{3m}=-m\frac{k_y}{k_{\perp}}\int_{b_1}^{\infty}x dx \cos k_x x\int_0^{\in
fty}u^2duJ_1(k_{\perp}u)R_2(r)/r^2\\
\Im I_{3m}= m\frac{k_y}{k_{\perp}}\int_{b_1}^{\infty}x dx \sin k_x x\int_0^{\in
fty}u^2duJ_1(k_{\perp}u)R_2(r)/r^2
\label{eq43}
\ea
For $l=2~m=\pm1$,
\be
A_{\delta 2m}=\frac{\sqrt{15}}{4\pi}(I_{1m}+I_{2})
\label{eq44}
\ee
with
,
\ba
I_{1m}=-m\frac{k_z}{k_{\perp}}\int_{b_1}^{\infty}dx e^{-ik_x x}\int_0^{\infty}u
^2duJ_1(k_{\perp}u)R_2(r)/r^2\\
\Re I_2=\frac{k_yk_z}{k_{\perp}^3}\int_{b_1}^{\infty} dx \sin k_x x\int_0^{\inf
ty}u^2duJ_1(k_{\perp}u)R_2(r)/r^2\nonumber\\
-\frac{k_yk_z}{k_{\perp}^2}\int_{b_1}^{\infty} dx \sin k_x x\int_0^{\infty}u^3d
u(J_0(k_{\perp}u)-J_2(k_{\perp}u))R_2(r)/r^2\\
\Im I_2=\frac{k_yk_z}{k_{\perp}^3}\int_{b_1}^{\infty} dx \cos k_x x\int_0^{\inf
ty}u^2duJ_1(k_{\perp}u)R_2(r)/r^2\nonumber\\
-\frac{k_yk_z}{k_{\perp}^2}\int_{b_1}^{\infty} dx \cos k_x x\int_0^{\infty}u^3d
u(J_0(k_{\perp}u)-J_2(k_{\perp}u))R_2(r)/r^2
\label{eq45}
\ea
Finally, for $l=2~m=0$ one has,
\be
A_{\delta 20}=\frac{\sqrt{10}}{8\pi}(3I_{1}+I_{2})
\ee
with
,
\ba
\Re I_1=-\frac{k_y^2}{k_{\perp}^3}\int_{b_1}^{\infty} dx \cos k_x x\int_0^{\inf
ty}u^2duJ_1(k_{\perp}u)R_2(r)/r^2\nonumber\\
-\frac{1}{2}\frac{k_z^2}{k_{\perp}^2}\int_{b_1}^{\infty} dx \cos k_x x\int_0^{\
infty}u^3du(J_0(k_{\perp}u)-J_2(k_{\perp}u))R_2(r)/r^2\\
\Im I_1=\frac{k_y^2}{k_{\perp}^3}\int_{b_1}^{\infty} dx \sin k_x
x\int_0^{\infty
}u^2duJ_1(k_{\perp}u)R_2(r)/r^2\nonumber\\
+\frac{1}{2}\frac{k_z^2}{k_{\perp}^2}\int_{b_1}^{\infty} dx \sin k_x x\int_0^{\
infty}u^3du(J_0(k_{\perp}u)-J_2(k_{\perp}u))R_2(r)/r^2\\
\Re I_2=\int_{b_1}^{\infty} dx \cos k_x x\int_0^{\infty}u^2duJ_0(k_{\perp}u)R_2
(r)/r^2\\
\Im I_2=-\int_{b_1}^{\infty} dx \sin k_x x\int_0^{\infty}u^2duJ_0(k_{\perp}u)R_
2(r)/r^2
\ea

\section{}\label{ppar_formulae}
In this appendix we give explicit formulae for the longitudinal momentum 
distributions
in the case of a planar cutoff. Complete transverse acceptances 
are assumed. The  momentum distribution is given by,
\be
\frac{dW}{d \vec k}=\frac{1}{\hat l^2}\sum_m \left| \int d\vec r
\frac{e^{-i\vec k\vec r}}{(2\pi)^{3/2}} \Phi_{lm}(\vec r) \right|^2
\ee
where $\Phi$ is a generic notation for $\delta\psi$ in the case of stripping or
$\psi_d$ in the case of diffraction. The differential cross section is then,
\be
\frac{d\sigma}{dk_z}=\int 2\pi b db\int_{-\infty}^{\infty} dk_x\int_{-\infty}
^{\infty} dk_y \frac{dW}{d \vec k}
\ee

$\bf{Stripping}$

$\bf{l=0~m=0}$
\be
\frac{d\sigma}{dk_z}=\frac{2}{\pi}\int_{b_{min}}^{b_{max}}b
db\int_{b_1}^{\infty}dx\int_0^{\infty}dy \left| \int_0^{\infty} dz 
\cos (k_zz)R_0(r) \right|^2
\ee 
$~~~~\bf{l=1~m=\pm 1}$
\be
\frac{d\sigma}{dk_z}=\frac{2}{\pi}\int_{b_{min}}^{b_{max}}b
db\int_{b_1}^{\infty}dx\int_0^{\infty}dy(x^2+y^2) \left| \int_0^{\infty} 
dz \cos (k_zz)R_1(r)/r \right|^2
\ee 
$~~~~\bf{l=1~m=0}$
\be
\frac{d\sigma}{dk_z}=\frac{2}{\pi}\int_{b_{min}}^{b_{max}}b
db\int_{b_1}^{\infty}dx\int_0^{\infty}dy \left| \int_0^{\infty} 
zdz \sin (k_zz)R_1(r)/r \right|^2
\ee 
$~~~~\bf{l=2~m=\pm 2}$
\be
\frac{d\sigma}{dk_z}=\frac{3}{2\pi}\int_{b_{min}}^{b_{max}}b
db\int_{b_1}^{\infty}dx\int_0^{\infty}dy(x^2+y^2)^2 \left| \int_0^{\infty} 
dz \cos (k_zz)R_2(r)/r^2 \right|^2
\ee 
$~~~~\bf{l=2~m=\pm 1}$
\be
\frac{d\sigma}{dk_z}=\frac{6}{\pi}\int_{b_{min}}^{b_{max}}b
db\int_{b_1}^{\infty}dx\int_0^{\infty}dy(x^2+y^2) \left| \int_0^{\infty} 
zdz \sin (k_zz)R_2(r)/r^2 \right|^2
\ee 
$~~~~\bf{l=2~m=0}$
\be
\frac{d\sigma}{dk_z}=\frac{1}{2\pi}\int_{b_{min}}^{b_{max}}b
db\int_{b_1}^{\infty}dx\int_0^{\infty}dy \left| \int_0^{\infty} dz 
\cos (k_zz)R_2(r)(3\frac{z^2}{r^2}-1) \right|^2
\ee 
Note that the apparent singularity in the above integrals is removed by the
behaviour of the wave function near origin $R_l(r)\sim r^{l+1}$. For diffraction 
the same expressions hold except that the integral over $x$ is
replaced by,
\be
\int_{-\infty}^{\infty}dx\vert f(x,b)\vert^2
\ee
where
\be
f(x,b)=1-\theta(x-b_1)-\gamma_{el}(b)e^{-iqx}
\ee
$b_1,\gamma_{el}$ and q are defined in the main text. $\theta$ is a step
function. $\frac{1}{\hat l^2}$ is already included in above expressions. The
total differential cross section is obtained simply by summing up contributions
from various $m$ values. For a given $l$ state, the longitudinal momentum
distribution is symmetric,
\be
\frac{d\sigma}{dk_z}(-k_z)=\frac{d\sigma}{dk_z}(k_z)  
\ee



\newpage

\onecolumn{  

\begin{figure}[h]
\begin{center}
\mbox{\epsfig{file=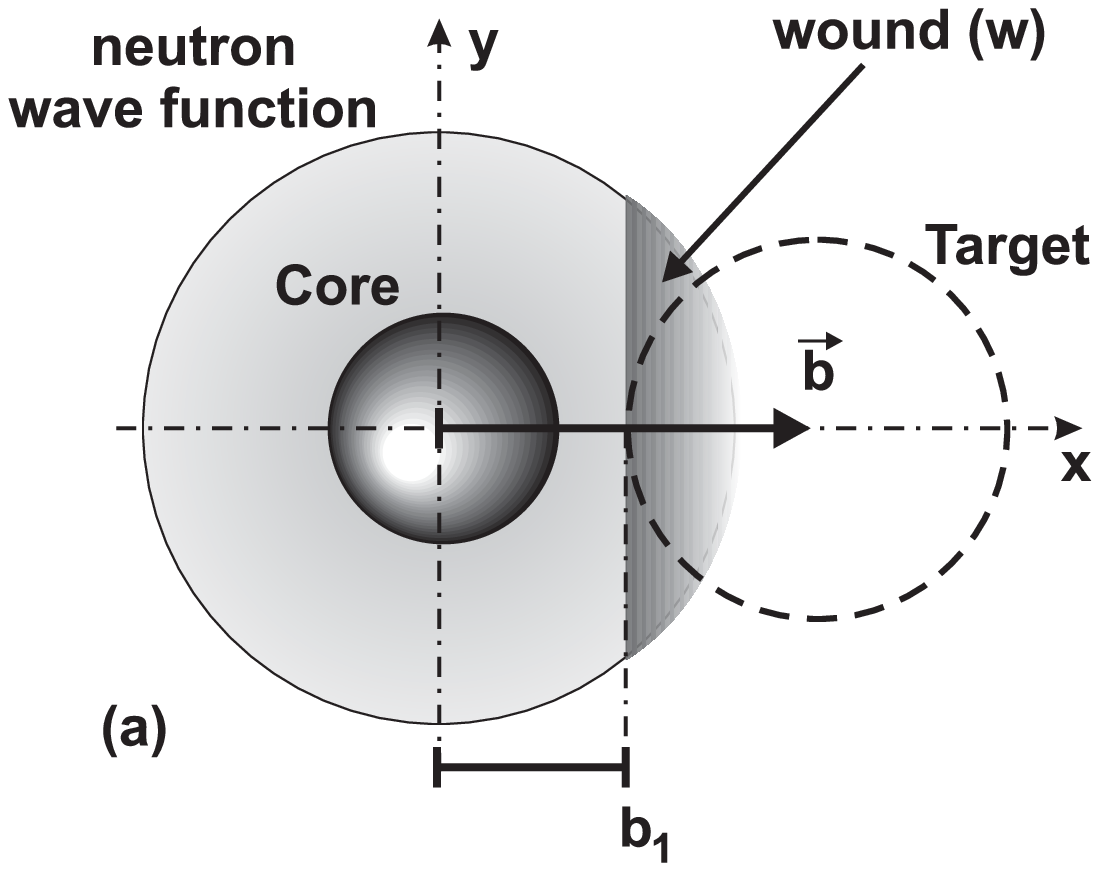,width=8cm}}
\mbox{\epsfig{file=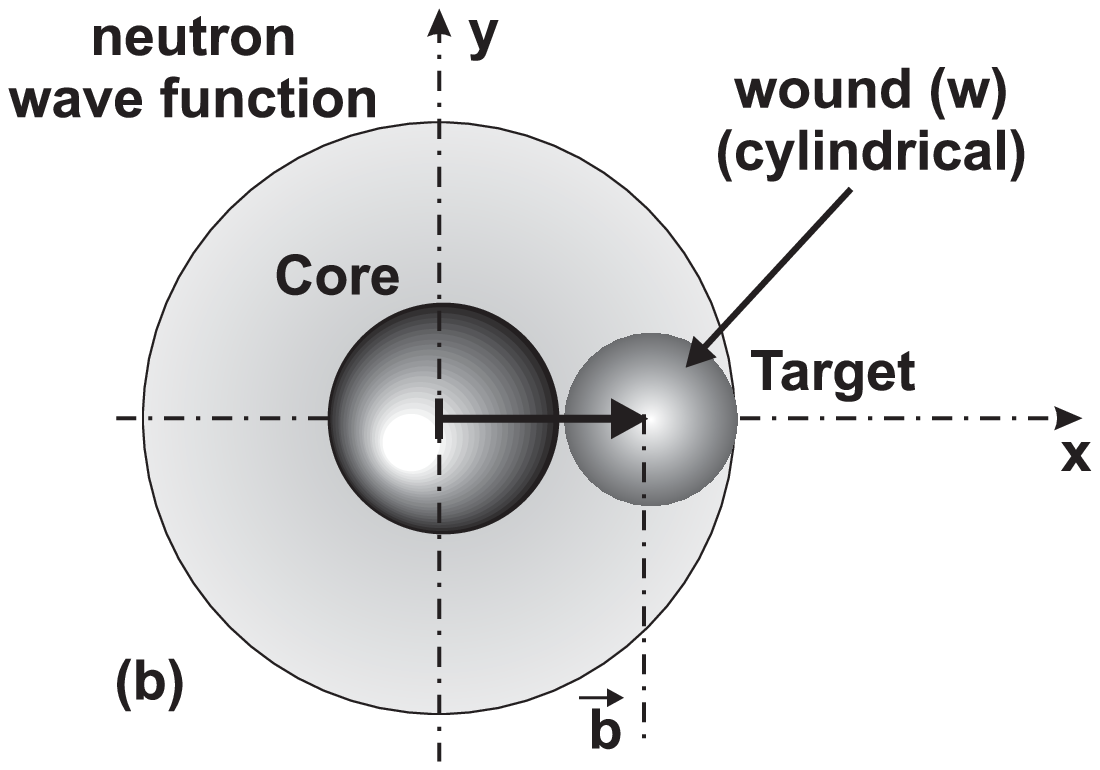,width=8cm}}
\end{center}
\caption{Schematic view of the ``wound'' induced by the projectile-target 
interaction for 
 (a) a planar cutoff and (b) a cylindrical wound. }
\label{fig:figsudd}
\end{figure}

\bigskip
\bigskip

\begin{figure}[h]
\begin{center}
\mbox{\epsfig{file=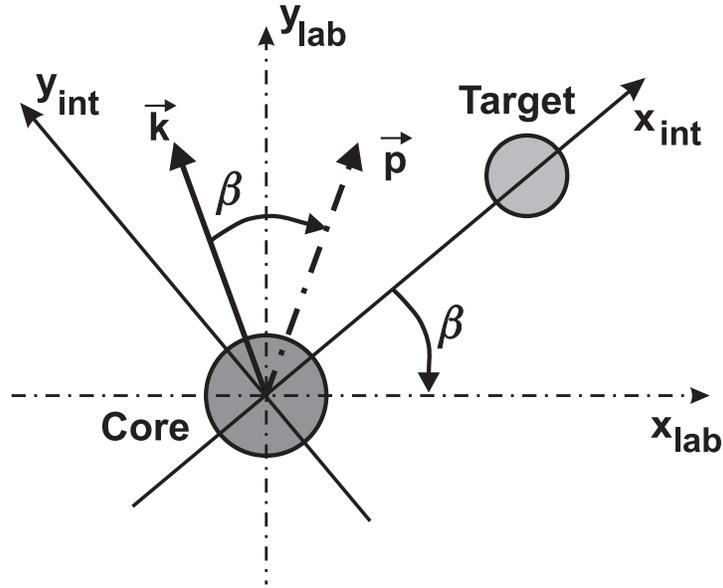,width=9.5cm}}
\end{center}
\caption{Schematic representation of the unitary transformation of Eq.~(\ref{eq47})
of the momentum 
vector 
$\vec k$ in the rest frame to the laboratory momentum ($\vec p$) by rotation 
through an angle $\beta$. }
\label{fig:figsudd2}
\end{figure}

\newpage

\begin{figure}[h]
\begin{center}
\mbox{\epsfig{file=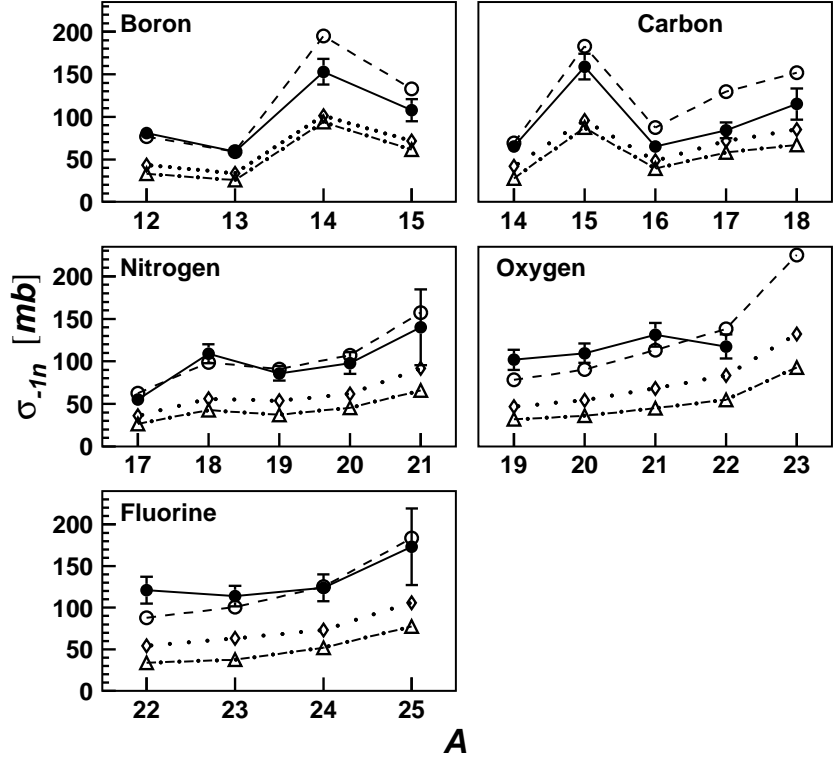,width=12.5cm}}
\end{center}
\caption{Single-neutron removal cross sections calculated in the planar cutoff
approximation
(open circles, dashed line) compared to the experimental values 
\protect\cite{nous,sauv01} 
(full circles, solid 
line) for reactions on a carbon target. The contributions arising from stripping 
(open triangles, dash-dotted line) and diffraction plus Coulomb dissociation 
(open diamonds, dotted line) are 
 also detailed.}
\label{fig:sefsud}
\end{figure}

\newpage

\begin{figure}[h]
\begin{center}
\mbox{\epsfig{file=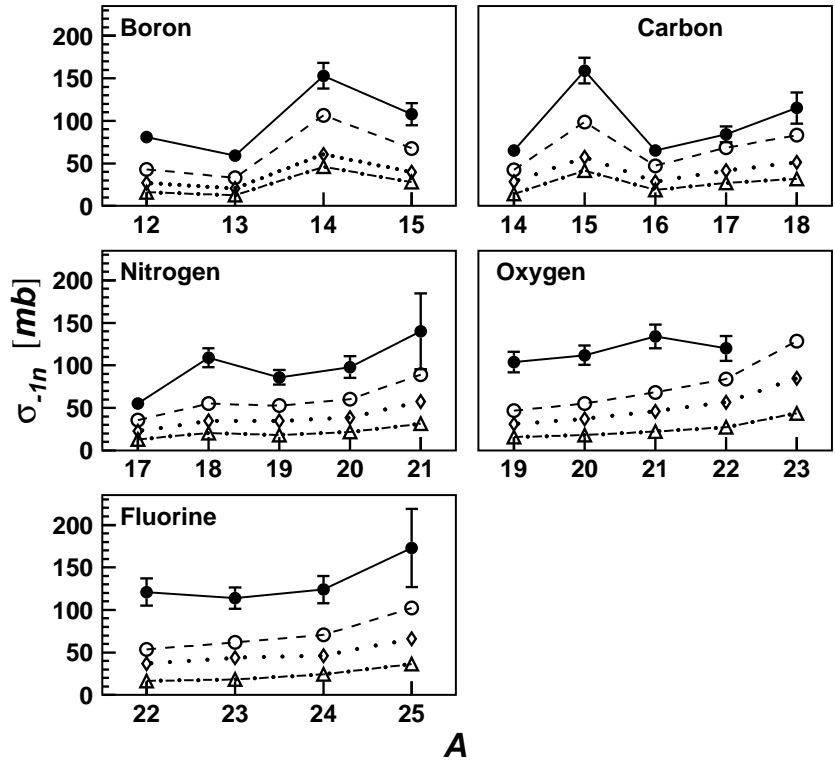,width=12.5cm}}
\end{center}
\caption{The same as in Fig.~\ref{fig:sefsud} but for a cylindrical wound. }
\label{fig:sefsud2}
\end{figure}

\newpage


\begin{figure}[!htbp]
\begin{center}
\mbox{\epsfig{file=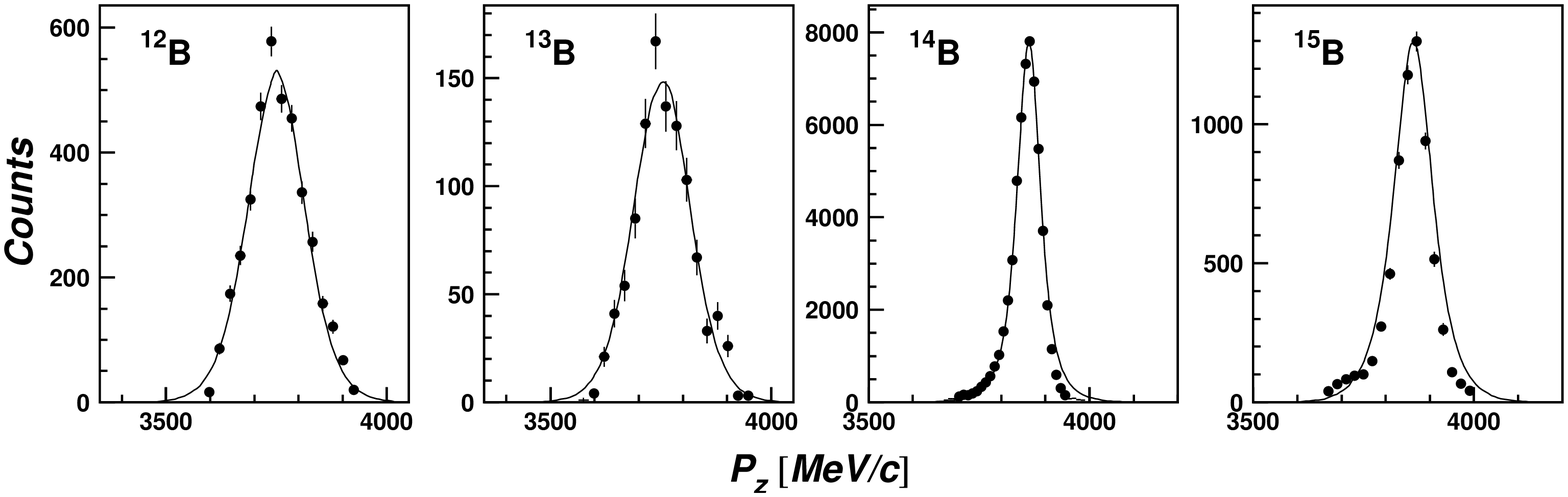,height=4cm}}
\mbox{\epsfig{file=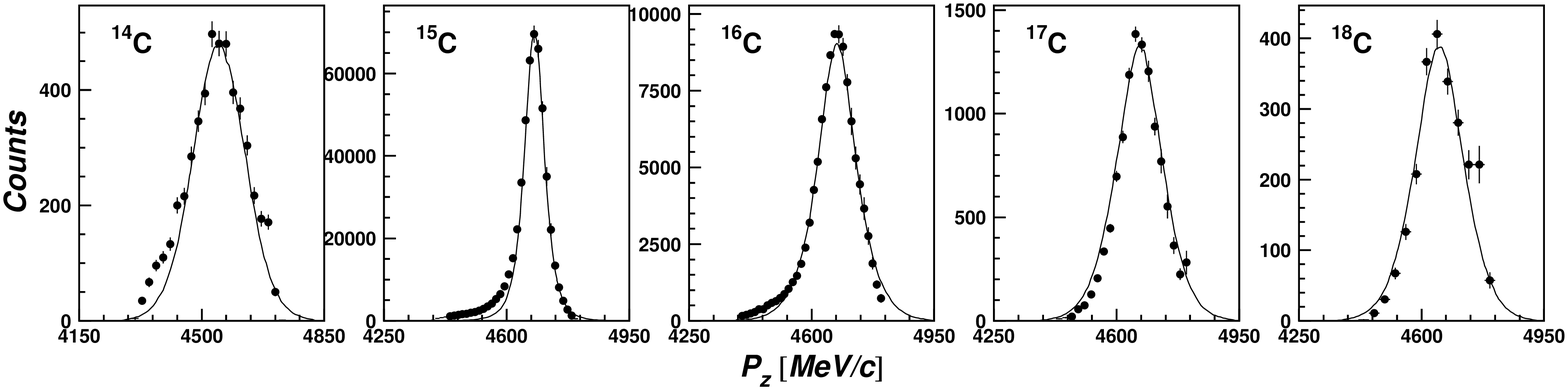,height=4cm}}
\mbox{\epsfig{file=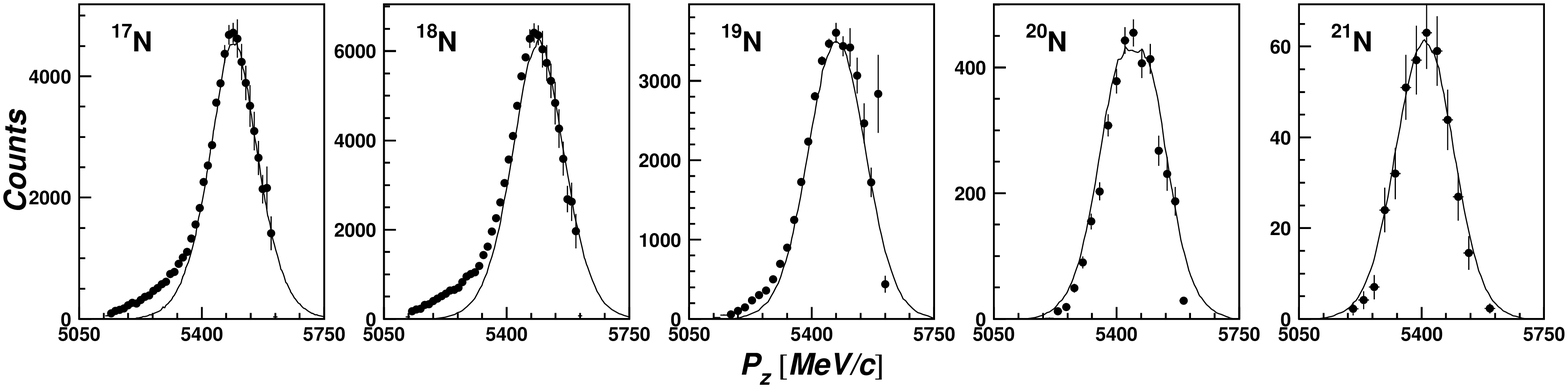,height=4cm}}
\mbox{\epsfig{file=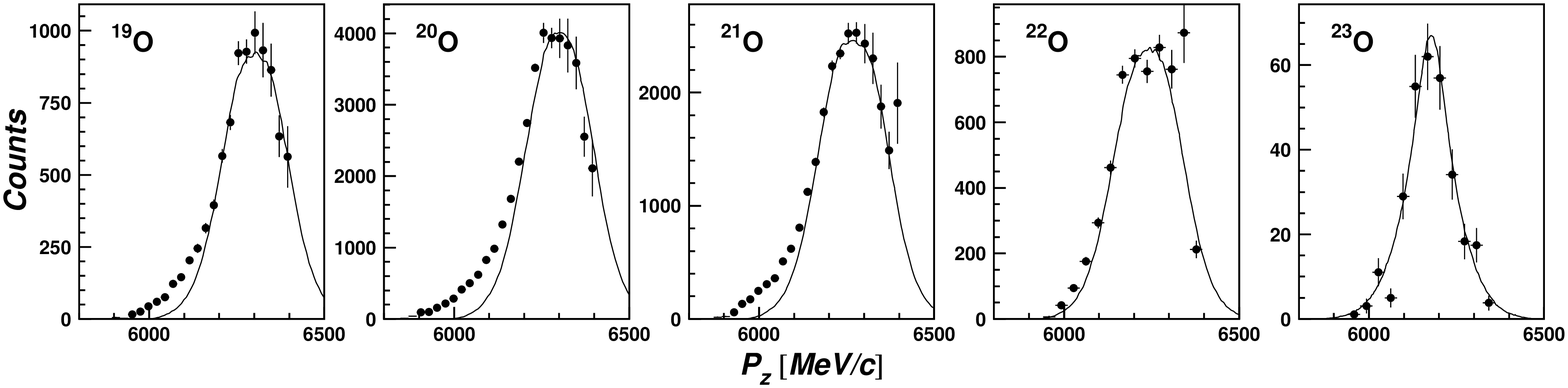,height=4cm}}
\mbox{\epsfig{file=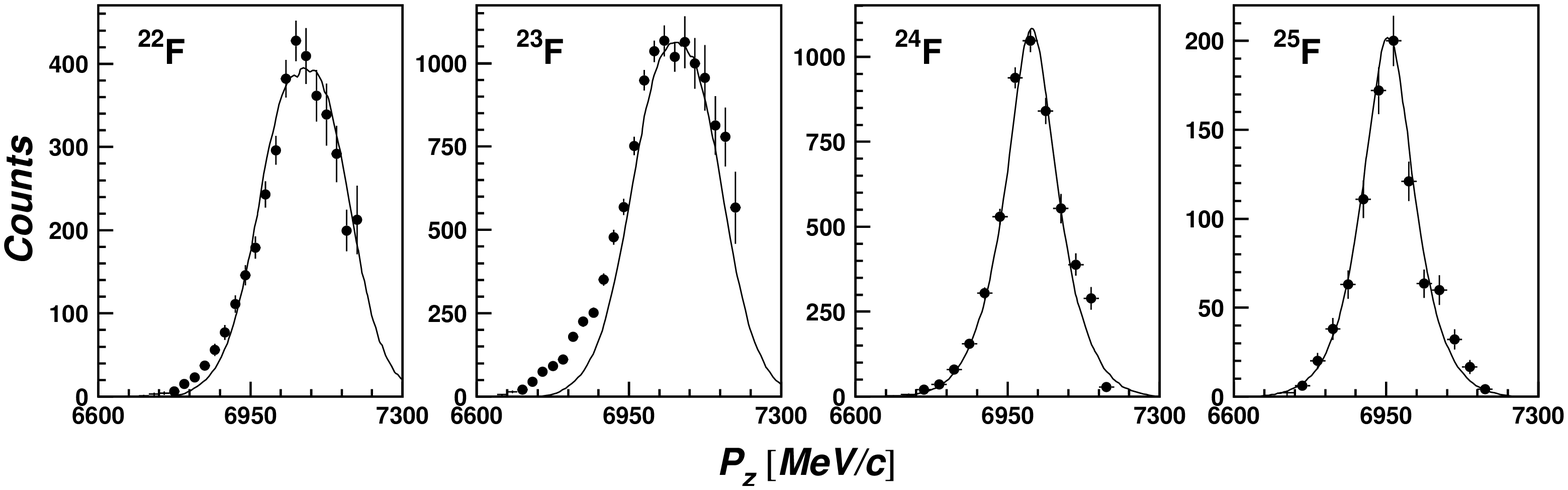,height=4cm}}
\end{center}
\caption{Core-fragment longitudinal momentum distributions for reactions 
on a carbon target. 
The experimental data are taken from ref. \protect\cite{nous,sauv01}.
The solid lines correspond 
to the results of calculations using the sudden approximation model and a 
planar cutoff.}
\label{fig:syssudd}
\end{figure}

\newpage

\begin{figure}[h]
\begin{center}
\mbox{\epsfig{file=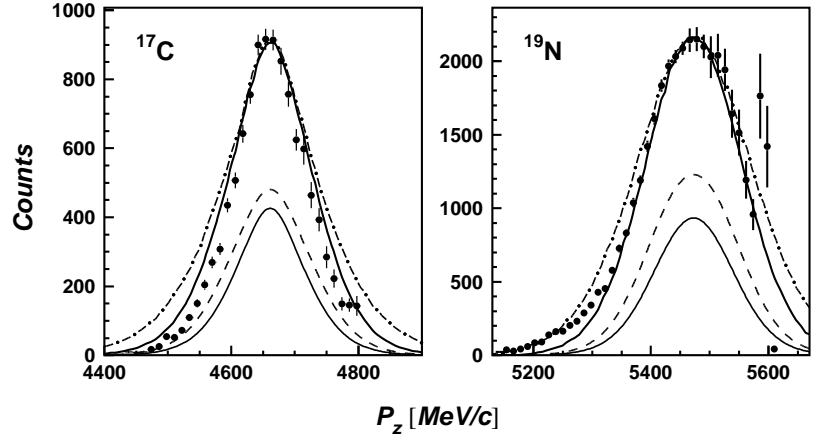,width=12.5cm}}
\end{center}
\caption{A comparison of the core-fragment longitudinal momentum distributions 
 calculated using the sudden model approximation with a planar cutoff 
 (solid line) and the 
Glauber model \protect\cite{sauv01} (dash-dotted line). The stripping and 
diffraction plus Coulomb components (thin solid and 
dashed lines respectively) in the sudden approximation are also shown.}
\label{fig:exsud}
\end{figure}

\newpage

\begin{figure}[h]
\begin{center}
\mbox{\epsfig{file=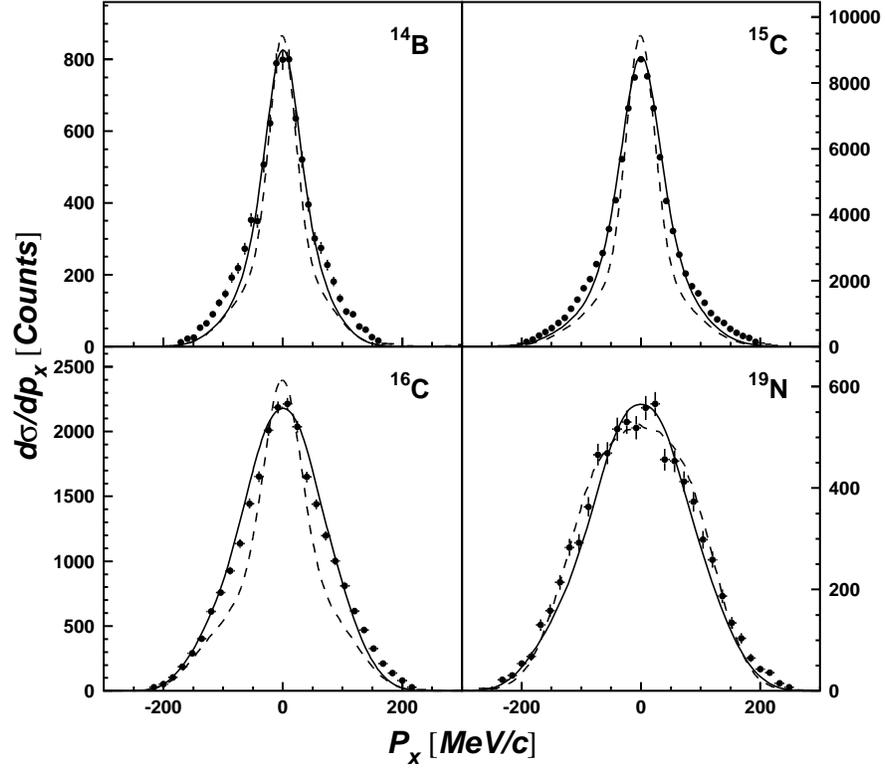,width=12.5cm}}
\end{center}
\caption{Comparison of the transverse (p$_x$) momentum distributions calculated 
within the 
sudden approximation (solid line)  with the Glauber model of 
ref. \protect\cite{sauv01} (dashed line). The  calculated distributions have
been convoluted 
with the experimental  effects.}
\label{fig:px_sudd_glau}
\end{figure}

\newpage

\begin{figure}[h]
\begin{center}
\mbox{\epsfig{file=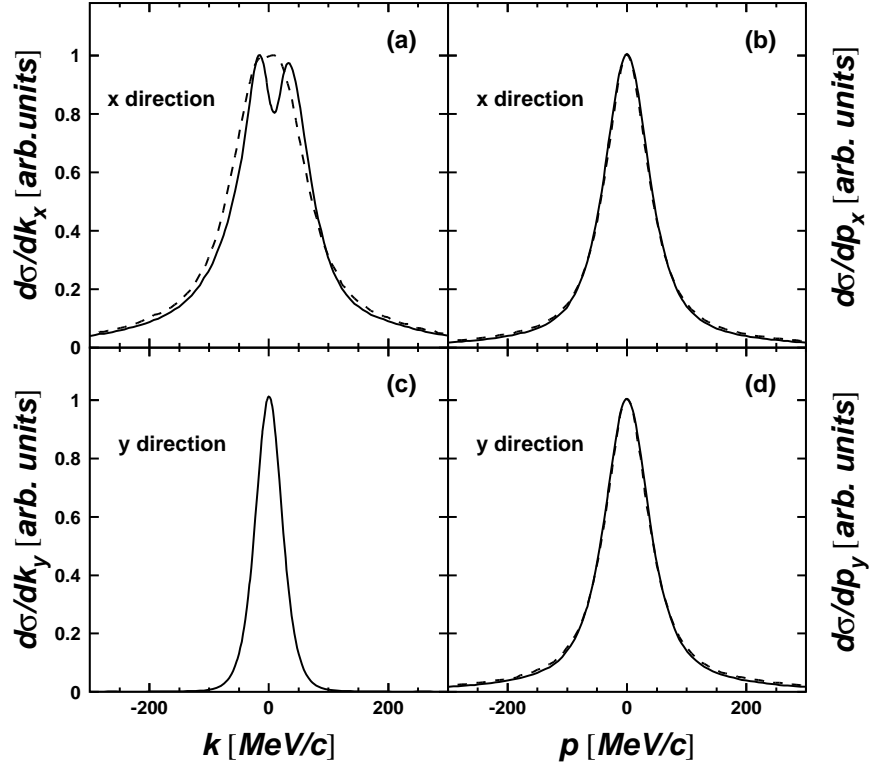,width=12.5cm}}
\end{center}
\caption{Momentum distribution in the projectile rest frame (panels a and c) and 
in the 
laboratory frame 
(panels b and d) for the $s$-wave ground state in $^{15}$C ($S_n=1.2$ MeV).
 Top panels: momentum distributions in the transverse $x$ direction. 
 Bottom panels: momentum distributions in the $y$ 
direction.  Calculations including Coulomb dissociation are shown by the solid
lines. Results without Coulomb interaction  are represented by dashed lines.
 }
\label{fig:s3d_kxpx}
\end{figure}

\newpage

\begin{figure}[h]
\begin{center}
\mbox{\epsfig{file=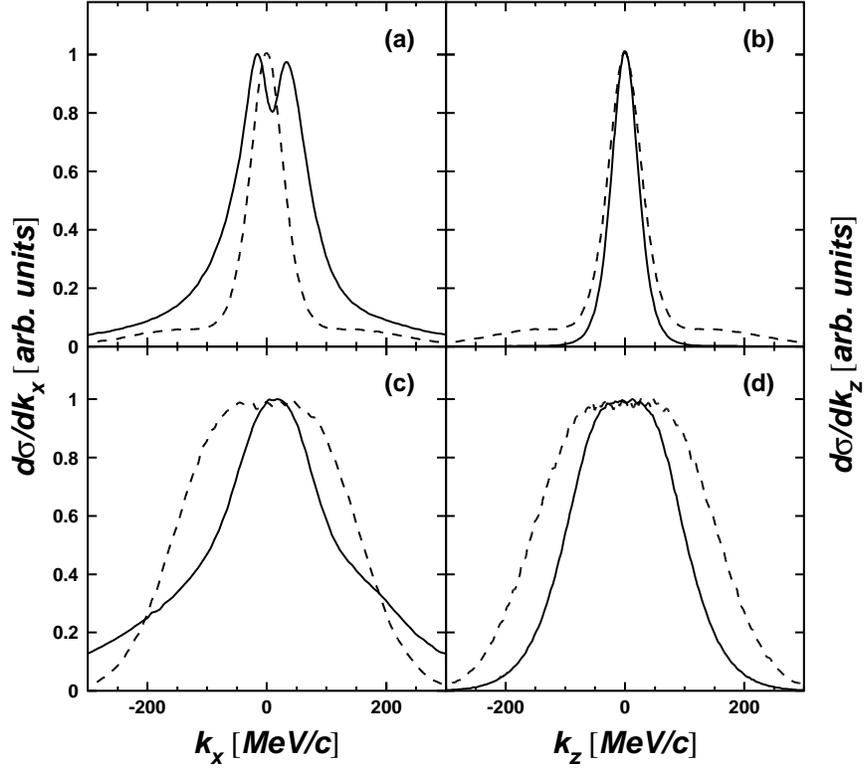,width=12.5cm}}
\end{center}
\caption{Intrinsic momentum distributions (dashed lines) compared with
the distributions from the reaction calculation for an $s$-wave valence neutron
in 
$^{15}$C (top panels) and a $d$-wave in $^{19}$N (bottom panels). Projections
in the transverse $x$ direction are shown in panels (a) and (c), whilst those 
in
longitudinal direction ($z$) are shown in (b) and (d).
}
\label{fig:s3d_int}
\end{figure}

\newpage

\begin{figure}[h]
\begin{center}
\mbox{\epsfig{file=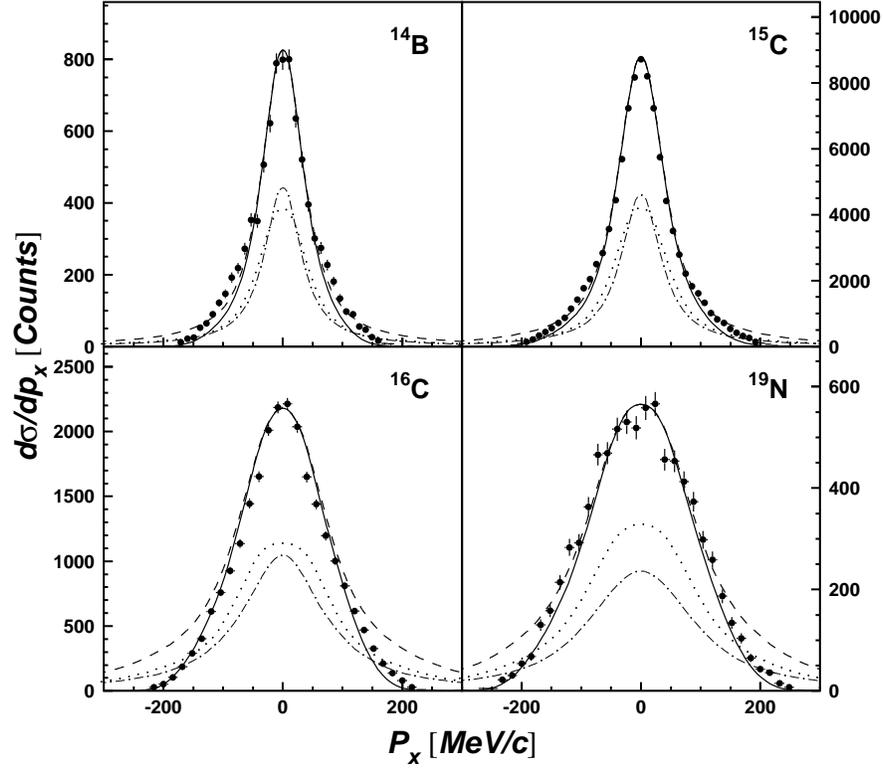,width=12.5cm}}
\end{center}
\caption{Comparison of the transverse (p$_x$) momentum distributions calculated 
within the 
sudden approximation with the data of ref. \protect\cite{sauv01} for $^{14}$B,
$^{15,16}$C and $^{19}$N. The decomposition of the total distribution 
 (dashed line)
   into stripping (dashed-dotted line) and diffraction plus Coulomb (dotted line) 
   components is shown. 
The calculated distributions, after Monte Carlo filtering to account for the 
experimental effects, are also displayed (solid line).}
\label{fig:px_sudd}
\end{figure}

\newpage

\begin{figure}[h]
\begin{center}
\mbox{\epsfig{file=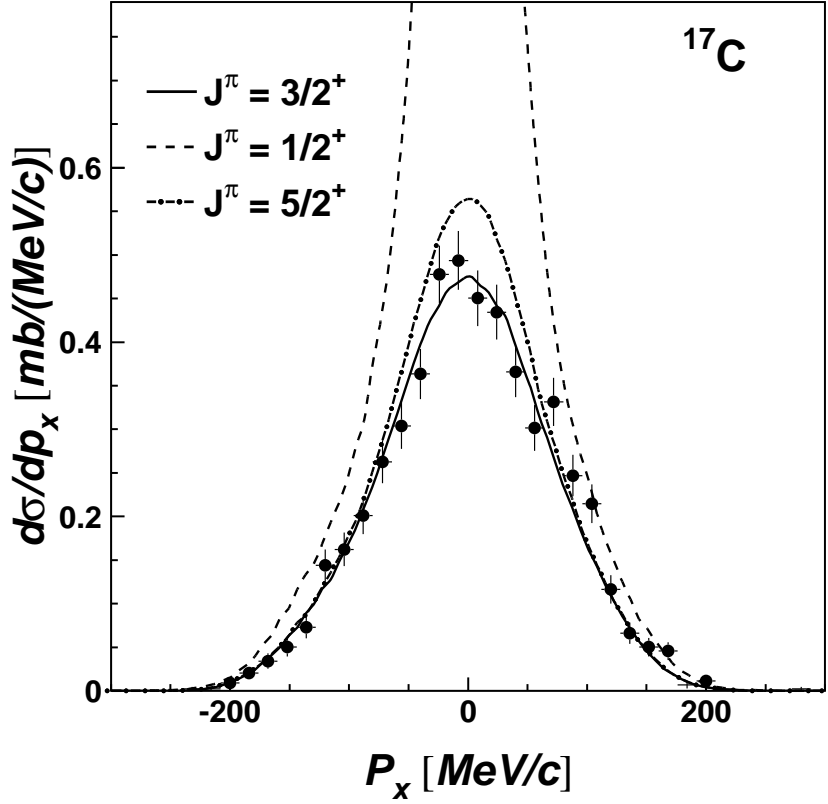,width=12.5cm}}
\end{center}
\caption{$^{16}$C core-fragment transverse momentum distributions from
reactions of $^{17}$C on a
carbon target (data ref. \protect\cite{sauv01})
calculated in 
the sudden approximation
for three spin-parity assignments $J^{\pi}$ for the ground state of $^{17}$C. 
}
\label{fig:spectro_px}
\end{figure}

\newpage

\begin{figure}[h]
\begin{center}
\mbox{\epsfig{file=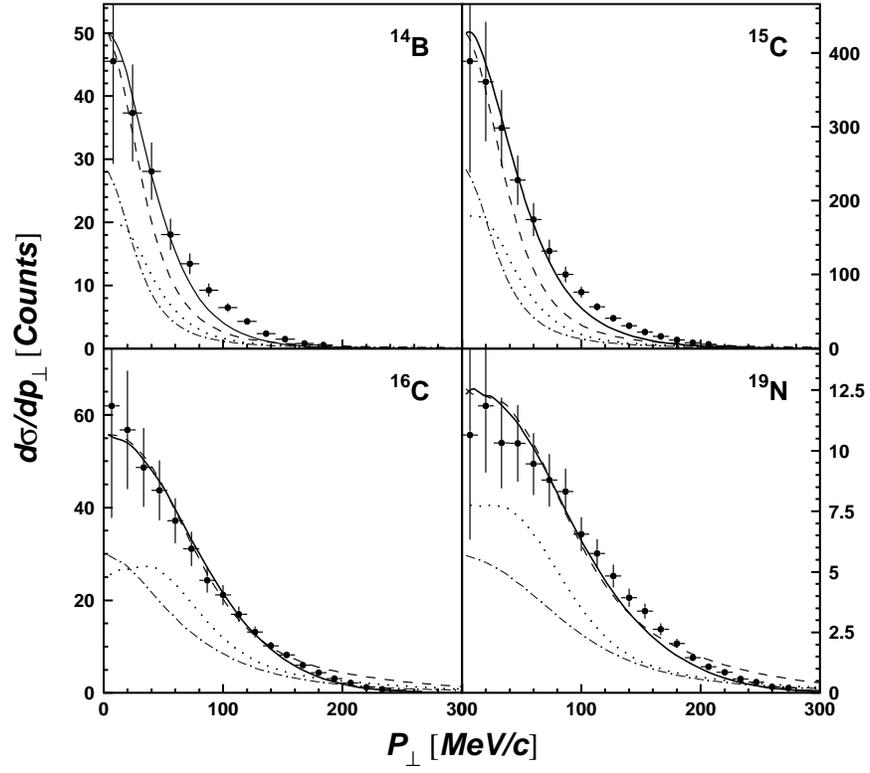,width=12.5cm}}
\end{center}
\caption{The same as in Fig.\protect\ref{fig:px_sudd} but for the perpendicular 
(p$_\perp$) momentum distributions calculated within 
the sudden approximation. }
\label{fig:pper_sudd}
\end{figure}

\newpage

\begin{figure}[h]
\begin{center}
\mbox{\epsfig{file=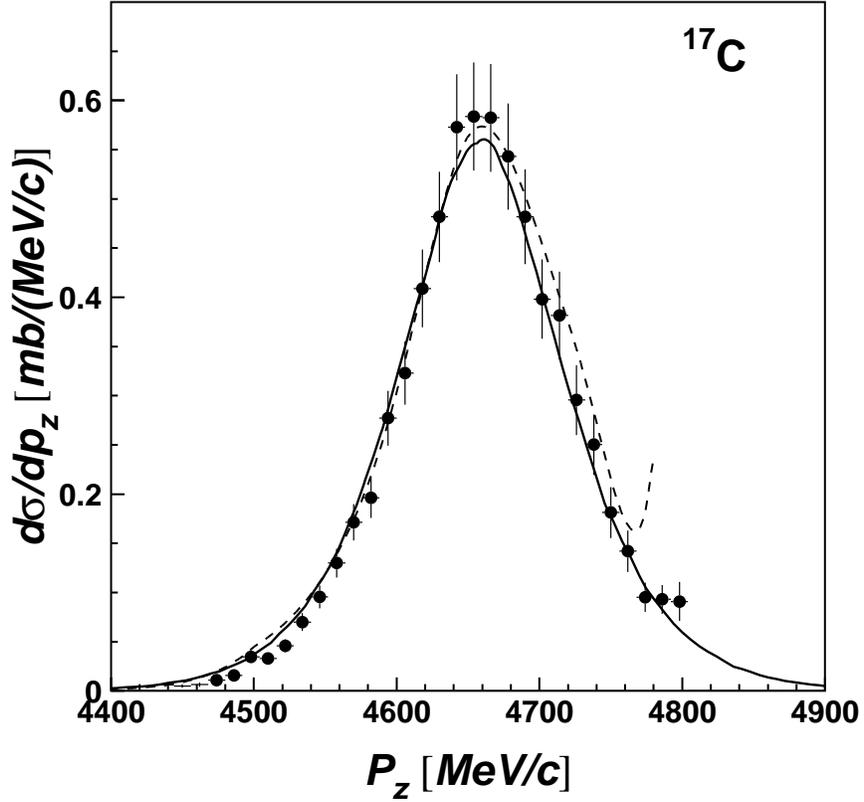,width=12.5cm}}
\end{center}
\caption{Comparison of the measured and predicted longitudinal momentum
distributions for single-neutron removal from $^{17}$C on a carbon target.
The results obtained using the sudden approximation model are shown by
the solid line, whilst those derived using the TCM (see text) are displayed by
the dashed line.}
\label{fig:TCM}
\end{figure}

\newpage

\begin{figure}[h]
\begin{center}
\mbox{\epsfig{file=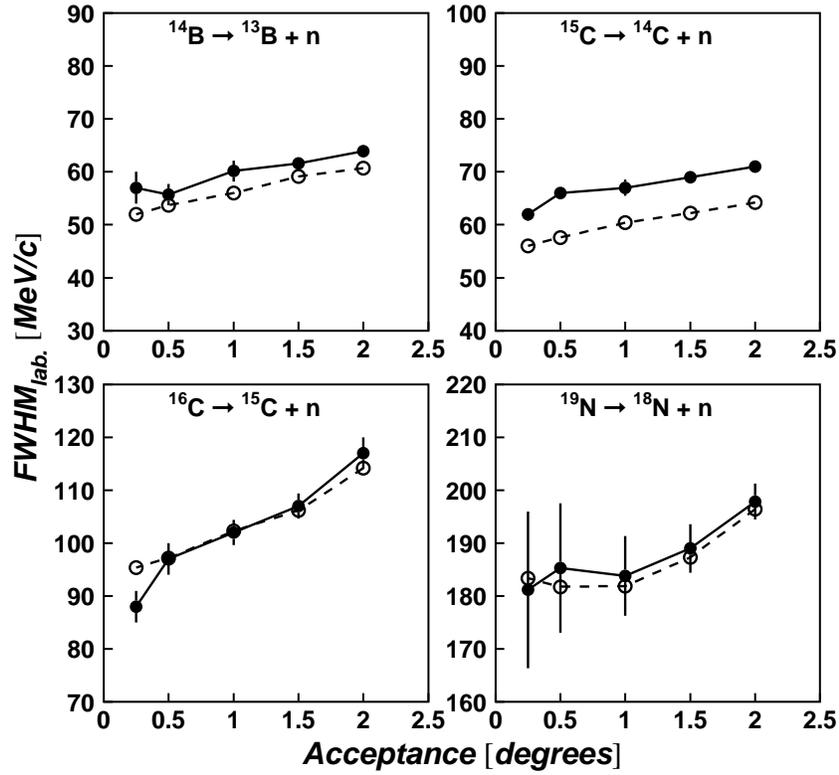,width=12.5cm}}
\end{center}
\caption{Evolution of the widths (FWHM) of the longitudinal momentum 
distributions 
with the  spectrometer angular acceptance. The experimentally observed trend 
 (filled circles, solid
line) is compared to the calculations in the sudden approximation model (open
circles, dashed lines). Note that the scale for the widths has been expanded.}
\label{fig:cut_angle_width}
\end{figure}

\newpage

\begin{figure}[h]
\begin{center}
\mbox{\epsfig{file=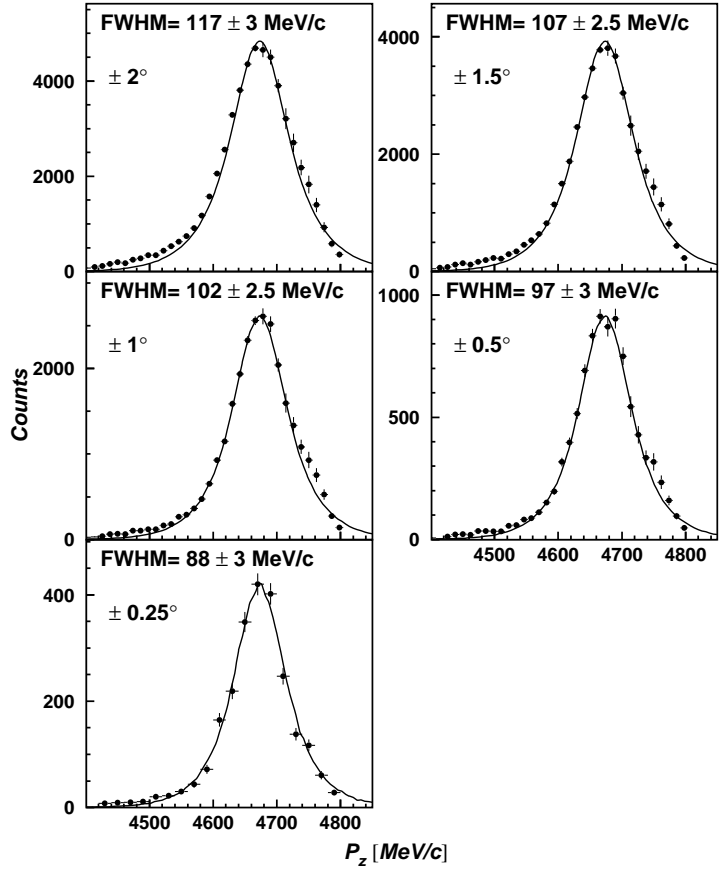,width=11cm}}
\end{center}
\caption{Evolution of the core-fragment longitudinal momentum distribution 
for reactions of $^{16}$C on carbon as a function of angular acceptances of the
spectrometer. The model calculations (solid lines) have been normalized to the
maximum number of counts and include, in addition to the
finite acceptances, 
all experimental broadening effects.}
\label{fig:cut_angle16C}
\end{figure}

\newpage

\begin{figure}[h]
\begin{center}
\mbox{\epsfig{file=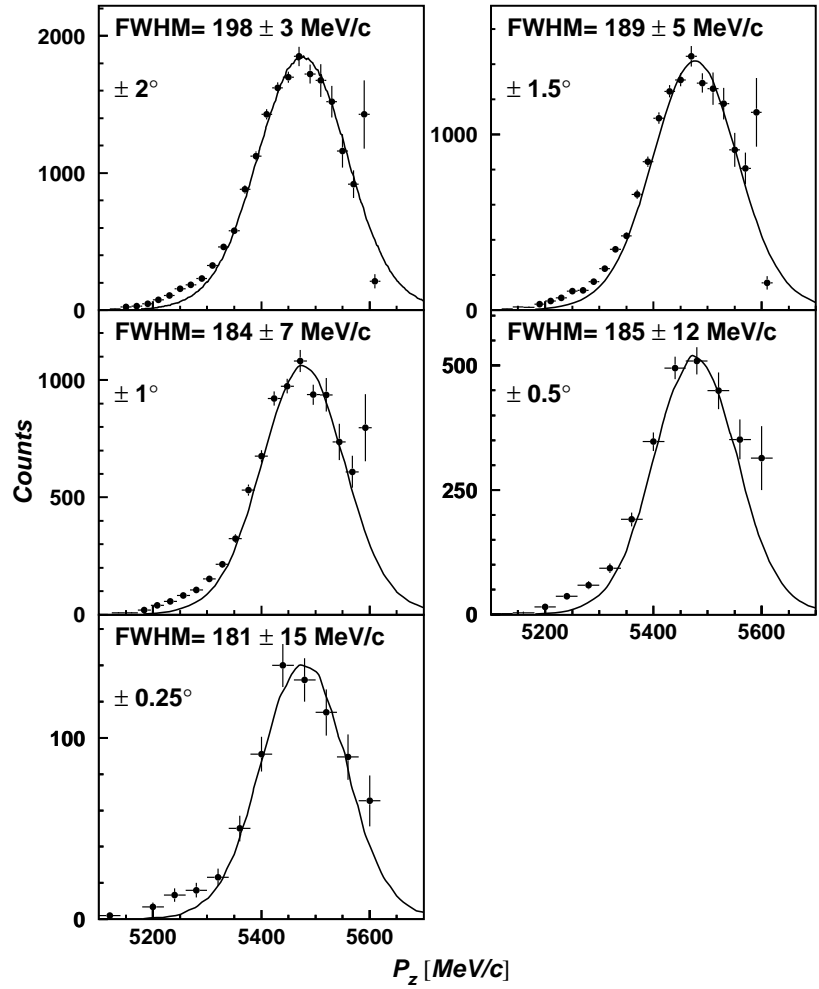,width=12.5cm}}
\end{center}
\caption{The same as in Fig.~\ref{fig:cut_angle16C} but for $^{19}$N.}
\label{fig:cut_angle19N}
\end{figure}


}

\end{document}